\documentclass{emulateapj}

\newcommand{\kms}{km~s$^{-1}$ }

\usepackage{lscape} 
\usepackage{color} 
\usepackage{graphics,graphicx}
\slugcomment{Resubmitted on July 17, 2015}
\shorttitle{Systematic Properties of Galactic Winds}
\shortauthors{Heckman et al.}

\begin{document}
\title{The Systematic Properties of the Warm Phase of Starburst-Driven Galactic Winds}
\author{Timothy M. Heckman}
\affil{Center for Astrophysical Sciences, Department of Physics \& Astronomy, Johns Hopkins University, Baltimore, MD 21218, USA and Institute of Astronomy, The University of Cambridge, Cambridge CB3 0HA, UK}
\author{Rachel M. Alexandroff} 
\affil{Center for Astrophysical Sciences,  Department of Physics \& Astronomy, Johns Hopkins University, Baltimore, MD 21218, USA}

\author{Sanchayeeta Borthakur }

\affil{Center for Astrophysical Sciences,  Department of Physics \& Astronomy, Johns Hopkins University, Baltimore, MD 21218, USA and The Department of Astronomy, The University of California, Berkeley, USA}
\author{Roderik Overzier}
\affil{Observatorio Nacional, Ministry of Science, Technology, and Innovation, Rio de Janeiro, Brazil}
\author{Claus Leitherer}
\affil{Space Telescope Science Institute, Baltimore, MD 21218, USA}

\begin{abstract}

Using ultra-violet absorption-lines, we analyze the systematic properties of the warm ionized phase of starburst-driven winds in a sample of 39 low-redshift objects that spans broad ranges in starburst and galaxy properties.  Total column densities for the outflows are $\sim$10$^{21}$ cm$^{-2}$. The outflow velocity (v$_{out}$) correlates only weakly with the galaxy stellar mass (M$_*$), or  circular velocity (v$_{cir}$), but strongly with both SFR and SFR/area.  The normalized outflow velocity (v$_{out}/v_{cir}$) correlates well with both SFR/area and SFR/M$_*$. The estimated outflow rates of warm ionized gas ($\dot{M}$) are $\sim$ 1 to 4 times the SFR, and the ratio $\dot{M}/SFR$  does not correlate with v$_{out}$.

We show that a model of a population of clouds accelerated by the combined forces of gravity and the momentum flux from the starburst matches the data. We find a threshold value for  the ratio of the momentum flux supplied by the starburst to the critical momentum flux needed for the wind to overcome gravity acting on the clouds ($R_{crit}$).  For $R_{crit} >$  10 (strong-outflows) the outflow's momentum flux is similar to the total momentum flux from the starburst and the outflow velocity exceeds the galaxy escape velocity. Neither is the case for the weak-outflows ($R_{crit} <$ 10). For the weak-outflows, the data severely disagree with many prescriptions in numerical simulations or semi-analytic models of galaxy evolution. The agreement is better for the strong-outflows, and we advocate the use of $R_{crit}$ to guide future prescriptions.

\end{abstract}
\keywords{ galaxies: starbursts --- galaxies: ISM ---galaxies:  intergalactic medium---galaxies: evolution---galaxies: kinematics and dynamics}

\section{INTRODUCTION}
In the standard picture, a galaxy will form when gas is able to reach high enough densities to cool, sink to the center of its dark matter halo, and form stars. What happens next depends on the interplay between a host of complex physical processes. In particular, massive stars produce a copious supply of ionizing radiation, and then quickly end their lives as supernovae. These processes ionize, heat, and chemically-enrich their surroundings. Unfortunately, a robust theoretical understanding of stellar feedback has proved elusive. The physical processes that control star formation and the resulting feedback operate on scales that are still well below the resolution of any cosmological simulation (`sub-grid physics'). These feedback processes are normally parametrized in both numerical simulations and semi-analytic models using prescriptions based on simple theoretical arguments and/or empirical relations (see Somerville \& Dav{\'e} 2015 for a recent review).

Galactic winds flowing from intensely star-forming galaxies are an especially important form of feedback. They are driven by the energy and/or momentum supplied by massive stars and their evolutionary descendants (e.g. Chevalier \& Clegg 1985; Murray et al. 2005). Winds like these are expected to alter the structure and evolution of the galaxy, possibly slowing down the rate at which it can subsequently form stars.  They can also expel metal-enriched material into the circum-galactic medium and even into the inter-galactic medium. These winds are complex multi-phase phenomena, with outflowing material spanning the range from cold molecular gas revealed in the mm and sub-mm regime to very hot gas studied in X-rays and the relativistic magnetized material seen in synchrotron emission in the radio (see Veilleux et al. 2005 and references therein). Thus, the full range of the wind properties and effects can only be studied in relatively nearby galaxies.

In the contemporary universe, strong winds are only observed in galaxies undergoing intense bursts of star formation. In contrast, at high redshift, winds are essentially ubiquitous in typical star-forming galaxies (e.g. Shapley 2011). Ideally then, the insights provided by detailed observations of winds in nearby starbursts should be combined with the demographic information about winds at high redshift, in order to understand the role of galactic winds in the evolution of galaxies and the inter-galactic medium.

The great majority of information about galactic winds at high redshift has come from observations of interstellar absorption lines in rest-frame ultraviolet (UV). These lines trace outflowing warm neutral and ionized gas (e.g. Shapley 2011; Steidel et al. 2010). \footnote{In this paper we use the term `wind' to describe the full multi-phase phenomenon. We use the term `hot wind fluid' to describe the outflowing thermalized ejecta from supernovae and stellar winds. A combination of this hot wind fluid and stellar radiation interacts with the interstellar or circum-galactic medium to produce the outflowing gas seen in absorption (which we will term the `outflow').}  They can also be observed in local starburst galaxies using space-based UV observations. The goal of the present paper is to use UV spectroscopic observations of a well-chosen and well-characterized sample of low-z starburst galaxies to determine which properties of the starburst and the surrounding galaxy correlate best with the principal properties of the outflows. Our hope is that these empirical correlations will not only lead to a better understanding of galactic winds, but can also be used to inform the sub-grid treatment of galactic winds in future numerical simulations.

In section 2 we describe the two samples of starbursts and summarize the analysis of the UV spectra. In section 3 we present empirical results on the correlations of measured outflow velocity and estimated mass and momentum outflow rates with the starburst and galaxy properties and compare our results with previous studies.  In section 4 we compare our observations with simple analytic models for momentum-driven winds and discuss the implications for models and simulations of galaxy evolution. We then summarize our results in section 5.  Throughout the paper we adopt a cosmological model with $H_0 =$ 70 km s$^{-1}$ Mpc$^{-1}$, $\Omega_{\Lambda} = 0.7$, and $\Omega_{Matter} = 0.3$.

\section{Sample Selection \& Data Analysis}
\subsection{The Samples}
We use two samples of low-redshift ($z < 0.2$) starburst galaxies in the analysis presented in this paper. The first is a sample of 19 galaxies observed with the Far Ultraviolet Spectroscopic Explore (FUSE), and was previously investigated by Grimes et al. (2009: hereafter G09) and Heckman et al. (2001).  The second is a sample of 21 Lyman Break Analogs (LBAs) observed with the Cosmic Origins Spectrograph (COS) on the Hubble Space Telescope (HST), and was previously investigated by Alexandroff et al. (2015: hereafter A15).  These papers contain descriptions of both the sample selection and the data analysis.  We refer the reader to these papers for the details, but as a matter of convenience we give a brief summary here.

The COS LBA sample was drawn from the large parent sample derived from a cross-match of GALEX GR6 and SDSS DR7 (Overzier et al. 2015).  The criteria used to define the parent sample were a K-corrected GALEX-based FUV luminosity $L_{FUV} > 10^{10.3}$ L$_{\odot}$ and a GALEX based FUV effective surface-brightness $\rm I_{FUV}  (=  1/2 L_{FUV}/ \pi  r_{50}^2) > 10^9~ L_{\odot} kpc^{-2}$, where $r_{50}$ was the seeing-deconvolved SDSS u-band half-light radius (Heckman et al. 2005; Hoopes et al. 2007).  These criteria were designed to select objects with luminosities and surface-brightnesses similar to typical Lyman Break Galaxies (LBGs) at $z \sim$ 3 to 4 (e.g. Shapley 2011). Subsequent observations have shown that the LBAs closely resemble LBGs in all salient properties: stellar mass, SFR, metallicity, galaxy size and morphology, gas kinematics, and gas and dust content (Hoopes et al. 2007; Basu-Zych et al. 2007; Overzier et al. 2010,2011; Gon\c{c}alves et al. 2010,2014). The specific targets observed with COS were selected based on maximizing the estimated FUV flux within the COS aperture.  These data include COS G130M and G160M spectra for each galaxy, and were obtained as part of the HST Cycle 17 Program 11727 and Cycle 20 Program 13017 (PI: T. Heckman).

The FUSE starburst sample was drawn from the archival collection of data on low-redshift star-forming galaxies. The specific members of this sample were selected for analysis solely on the basis of data quality (S/N $ >$  4 in 0.078~\AA\ bins in the galaxy continuum in the LiF 1A FUSE band).  These galaxies are broadly representative of the local population of UV-bright starburst galaxies. Three members of the FUSE sample (VV 114, Haro 11, and Mrk 54) meet the LBA criteria (Grimes et al. 2006,2007; Hoopes et al. 2007).

We have selected these two samples for several reasons. First, the spectroscopic aperture used is large enough to encompass the majority of the far-ultraviolet light emitted by the starburst. More quantitatively, the size of the COS aperture exceeds the UV half-light diameters (see below) of the galaxies by an average factor of 8 and by a minimum of factor of 3. The size of the FUSE aperture exceeds the UV half-light diameters (see below) by an average factor of 4.4. Only in the case of M 83 is the FUSE aperture smaller than the galaxy UV half-light diameter. This insures that we are measuring the global properties of the starburst in nearly every case.

Secondly, these two data sets are complementary.  As noted above, the COS LBA sample is representative of typical star-forming galaxies at high redshift. The FUSE starburst sample extends to galaxies of lower stellar mass, star-formation rates, and metallicity. This is important, since the goal in this paper is to quantify the dependence of the wind properties on these basic parameters. Third, both the FUSE and COS data have point source spectral resolutions (R $\sim$ 21,000 and 18,000 respectively). Even for extended targets, the effective resolution is sufficient to resolve the interstellar absorption-lines in these galaxies (unlike IUE, HUT, or HST/STIS).  It is also important to note that we have undertaken the analysis of both sets of data and have been uniform in the way we have measured the properties of the interstellar absorption-lines used to probe galactic winds.

This sample complements previous similar investigations. With the exception of the optically-selected sample in Chen et al. (2010), the previous analyses of comparably large samples of low-redshift galaxies have focused on dusty far-IR-selected galaxies (e.g. Heckman et al. 2000; Rupke et al. 2005; Martin 2005; Sato et al. 2009). All these low-z studies have used the optical Na I D doublet to probe the outflows. Our sample is essentially UV-selected, making it a better match to typical high-z star-forming galaxies (e.g. LBGs), and giving us access to diagnostic interstellar lines in the far-UV spanning a range in ionization state and optical depth.

There have also been a number of investigations of intermediate redshift ($z \sim$ 0.5 to 1.5) star-forming galaxies (e.g. Weiner et al. 2009; Erb et al. 2012; Kornei et al. 2012; Martin 2012; Rubin et al. 2014; Bordoloi et al. 2014). These have typically used stacked spectra of the Mg II 2800 doublet and/or several near-UV Fe II lines. Our sample spans a much larger range in star-formation rate and galaxy mass, giving us better leverage on the influence these properties have on those of the outflows. We also have adequate signal-to-noise and spectral resolution to be able to characterize the outflows in individual galaxies.

The principal properties of our merged COS and FUSE sample are listed in Table 1.  The starburst half-light radii (r$_*$) for the LBAs were measured using HST UV images (A15), and from the available UV images described in G09 for the FUSE sample.  The stellar masses (M$_*$) for the LBAs were taken from the MPA-JHU galaxy catalog for SDSS DR7. \footnote{Available from http://www.mpa-garching.mpg.de/SDSS/DR7/} The stellar masses for the FUSE sample were based on 2-MASS K-band luminosities and the K-band  mass/light ratio for strongly star-forming galaxies derived by Bell \& De Jong (2001).  Star-formation rates (SFR) were derived using a combination of GALEX FUV, IRAS and Herschel far-IR, and WISE and Spitzer mid-IR photometry following the prescriptions in Kennicutt \& Evans (2012), as described further in A15.  In all cases, a Kroupa/Chabrier Initial Mass Function was adopted for masses and SFRs. We have then defined an effective star-formation rate per unit area as $SFR/2\pi  r_*^{2}$. The metallicities (O/H) are based on the O3N2 strong-line estimator derived by Pettini \& Pagel (2004), using SDSS spectra for the LBA sample, and literature spectra for the FUSE sample.

We have determined a characteristic internal velocity parameter for each galaxy. Gon\c{c}alves et al. (2010) showed that both large-scale rotation and turbulent motions were significant in the kinematics of the interstellar medium of LBAs, and this is also seen in more typical local starburst galaxies (e.g. Lehnert \& Heckman 1996). We therefore use the kinematic parameter $S$ (Weiner et al. 2006; Kasim et al. 2007)  to characterize each galaxy: $S^2 = 0.5 v_{rot}^2 + \sigma_{gas}^2$  (where $v_{rot}$ and $\sigma_{gas}$ represent the observed rotation velocity and velocity dispersion of the ISM).  We adopt a value based on the good fit between $S$ and M$_*$ for low-redshift star-forming galaxies shown in Simons et al. (2015): $\rm log~S = 0.29~log~M_* - 0.93$, where $\rm S$ is in km sec$^{-1}$, and $M_*$ is in solar masses. We find that this relationship is a good fit to the data for the 16 LBAs in Gon\c{c}alves et al. (of which 6 are in the COS sample): the mean residual of the (observed minus fit) value for $log~S$ is -0.01 dex, with a standard deviation of 0.10 dex. In the rest of the paper we use the value of $S$ from this fit to estimate an effective circular velocity for each galaxy: v$_{cir} = \sqrt{2} S$ (where the $\sqrt{2}$ factor insures that $v_{cir} = v_{rot}$ in the limit $v_{rot} >> \sigma_{gas}$). \footnote{We also estimated values for the halo virial velocity ($v_{vir}$) using the scaling relation between $M_*$ and halo virial mass from Behroozi et al. (2010). Over the range in stellar mass where their relationship is calibrated ($log~M_* \geq$ 9.0), the values for $v_{vir}$ agree with our estimates of $v_{cir}$ to within 0.1 dex. We use $v_{cir}$ in this paper because its calibration has been more directly determined from data for strongly star-forming galaxies, and because it is calibrated over the full range of $M_*$ we are probing.}

\subsection{Analysis of the UV Spectra}
We refer the reader to G09 and A15 for detailed discussions of the data reduction for the FUSE and COS data respectively. In this paper we use the reduced spectra reported in these papers to measure two quantities: the column densities and velocity of the outflow based on UV interstellar absorption-lines.

\subsubsection{Column Densities}

The strong interstellar lines in these spectra are saturated (highly optically-thick) and therefore cannot be used to derive reliable ionic column densities.  Instead, we will use optically-thin (and therefore weak) lines.  In order to measure these weak features in the COS sample, we have created a high signal-to-noise stacked spectrum.  The procedure is discussed in A15 and the resulting stack is shown that paper.  From this stack we have used the following weak lines to trace both the neutral and ionized gas: S III 1012, O I 1039; S IV 1063, Fe II 1145, 1608; Al II 1670. The FUSE data are higher in signal-to-noise, so that the following weak lines were measured by G09 for individual galaxies: S III 1012, Si II 1021, OI 1039, S IV 1063, and Fe II 1097, 1145.

Ionic column densities were obtained by integrating the apparent optical depths of these lines over the velocity range covered by the absorption (Sembach et al. 2003; G09), using oscillator strengths from the on-line NIST Atomic Spectra Database. \footnote{http://www.nist.gov/pml/data/asd.cfm}  These column densities can be used to estimate total column densities for the neutral and warm ionized gas. The neutral phase is traced most naturally by O I. However,  S II, Si II, Fe II, and Al II are the dominant ionic species of these elements in the warm neutral atomic (HI) phase of the ISM in the starbursts since they all have ionization potentials less than a Rydberg.  We use the sum of the S III and S IV column densities to trace the ionized gas.

For the COS stacked spectrum we measured  the following column densities (cm$\rm ^{-2}): ~N_{O I}~<~10^{15.7},~N_{Fe II}~=~10^{14.6},~ N_{Al II}~=~10^{13.4}$. We have converted these into HI column densities using the mean O/H abundance derived from the nebular emission-lines for the sample (0.5 solar) and assuming solar O/Fe and O/Al abundance ratios.  These yield values for N$_{H I}$ of $< \rm 10^{19.3}$, $\rm 10^{19.4}$, and $\rm 10^{19.3}$~cm$^{-2}$  based on the O I, Fe II, and Al II lines respectively. The column density derived from O I is most reliable (least sensitive to dust depletion or ionization corrections). The estimated column densities are consistent with the lack of damping wings on the Ly$\alpha$ absorption-lines in the COS data for our LBAs.  The same exercise yields column densities of $\rm N_{S III} =  10^{15.3}$ cm$^{-2}$ and $\rm N_{S IV} = 10^{15.2}$~cm$^{-2}$. Sulfur is not significantly depleted onto dust grains, and we do not expect a significant amount of S V. The combined column density of these two ions corresponds to  $\rm N_{H II} = 10^{20.7}~cm^{-2}$.

For the FUSE data we can use the same approach but derive column densities for individual galaxies, converting ionic to Hydrogen column densities using the metallicities for each galaxy listed in Table 1. We see no evidence for significant systematic variations in H column densities from galaxy to galaxy,  so we simply calculated the mean values for the FUSE sample:  N$_{H I} = 10^{19.6}$ and N$_{H II} = 10^{20.6}$ cm$^{-2}$.  These are similar to the values for the COS sample.

\begin{figure}
\begin{center}
\includegraphics[trim=0mm 20mm 0mm 130mm,  clip=true,scale=0.45]{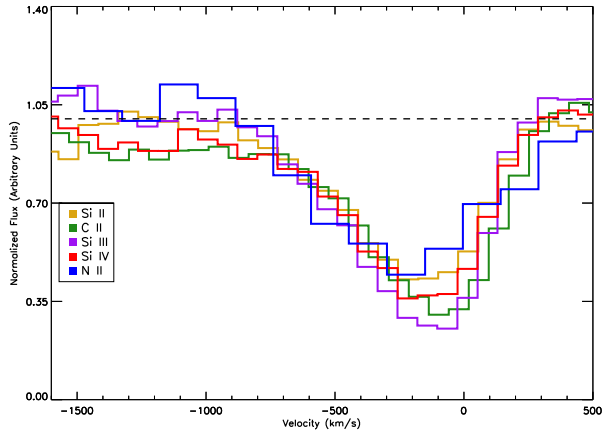}
\end{center}
\caption{We compare the absorption-line profiles of five strong interstellar transitions, extracted from the stacked, smoothed ($\sim$ 75 km s$^{-1}$ FWHM) high signal-to-noise spectrum of the COS sample described in Alexandroff et al. (2015). The continuum was normalized to unit flux locally for each of the lines in the figure. The figure shows that the profile shape of the Si III 1206 line that we use to measure outflow velocities in the COS sample (shown in purple) is very representative of the profile of the N II 1084 line (in blue) that we use for the FUSE sample, and to the profiles of the other strong lines (Si II 1260 – in yellow, C II 1334 – in green, and Si IV 1393 – in orange). The velocities on the x-axis are measured relative to the galaxy systemic velocity.} 
\end{figure}

\begin{figure*}
\begin{center}
\includegraphics[trim=0mm 0mm 33mm 0mm,  clip=true,scale=0.55]{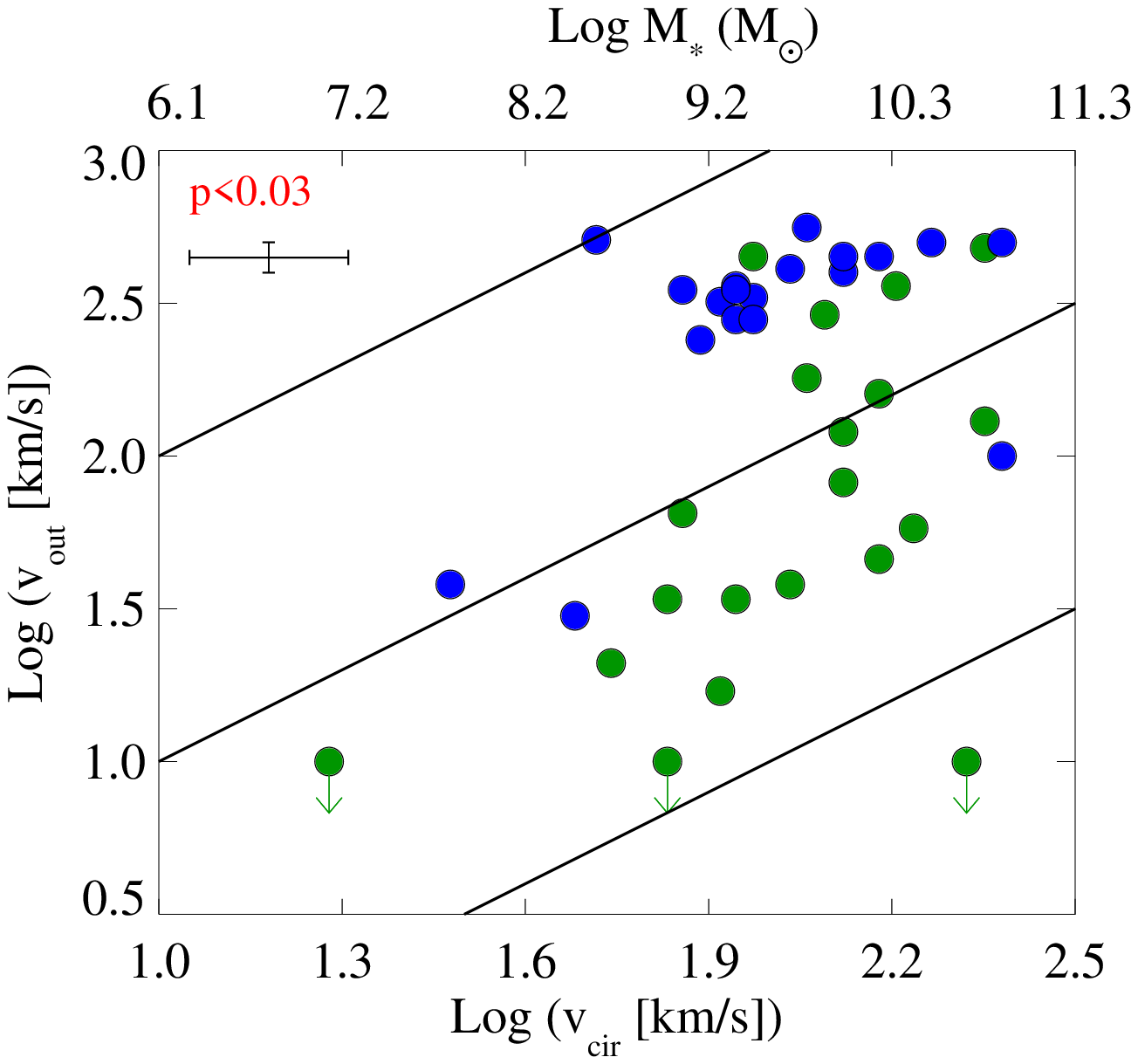}
\includegraphics[trim=0mm 0mm 33mm 0mm,  clip=true,scale=0.55]{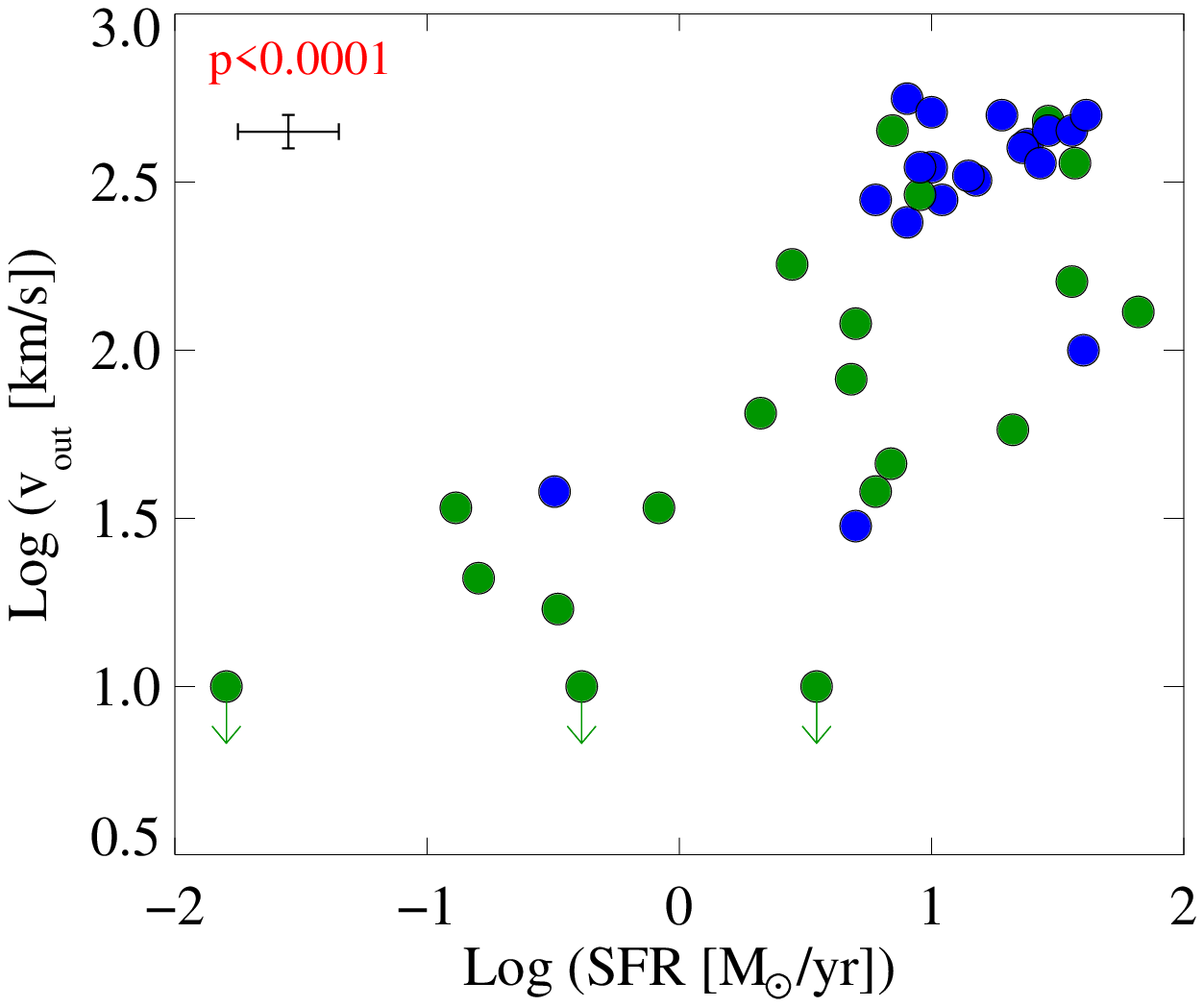}\\

\hspace{0.5cm}
\includegraphics[trim=0mm 0mm 33mm 0mm,  clip=true,scale=0.55]{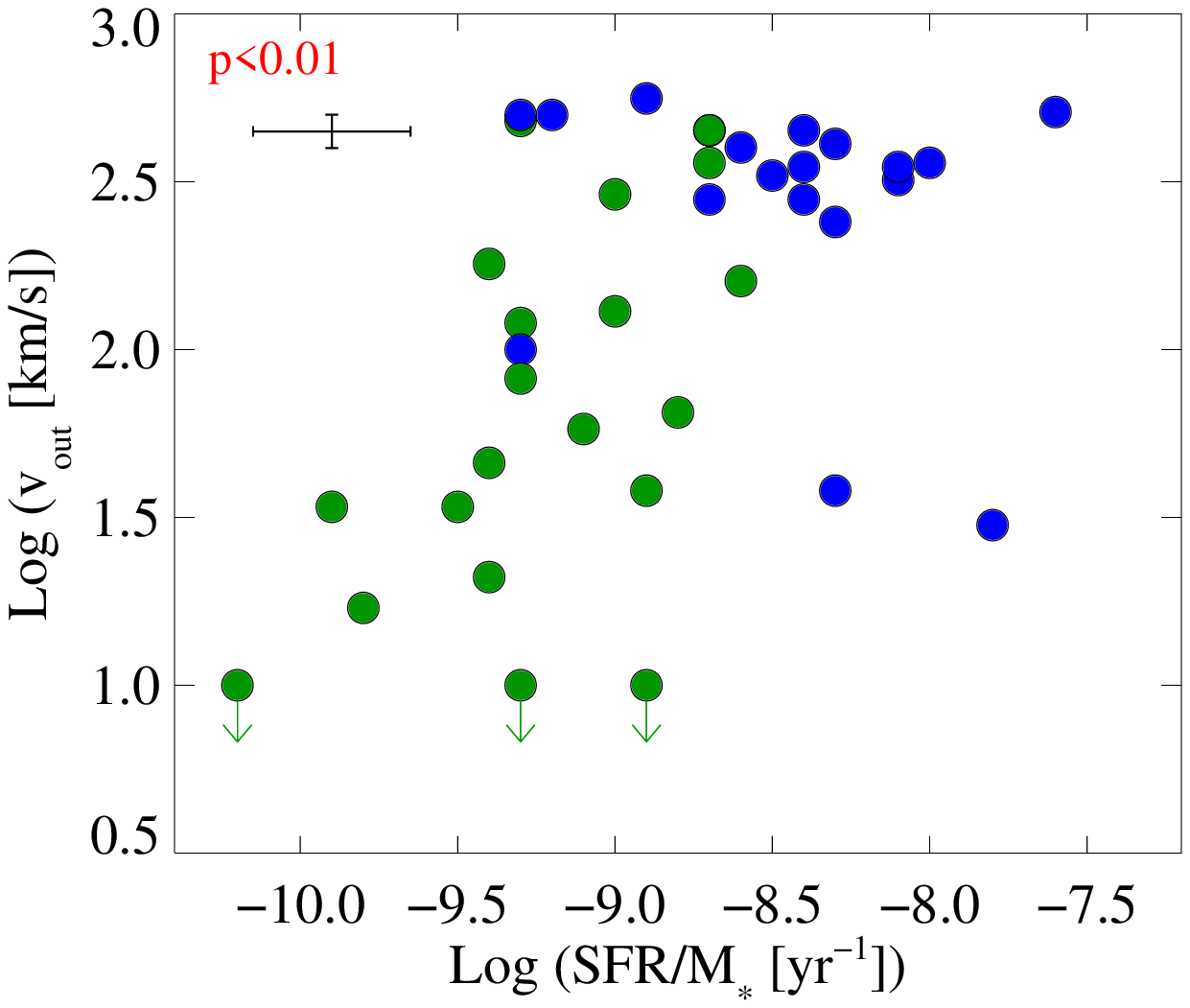}
\includegraphics[trim=0mm 0mm 33mm 0mm,  clip=true,scale=0.55]{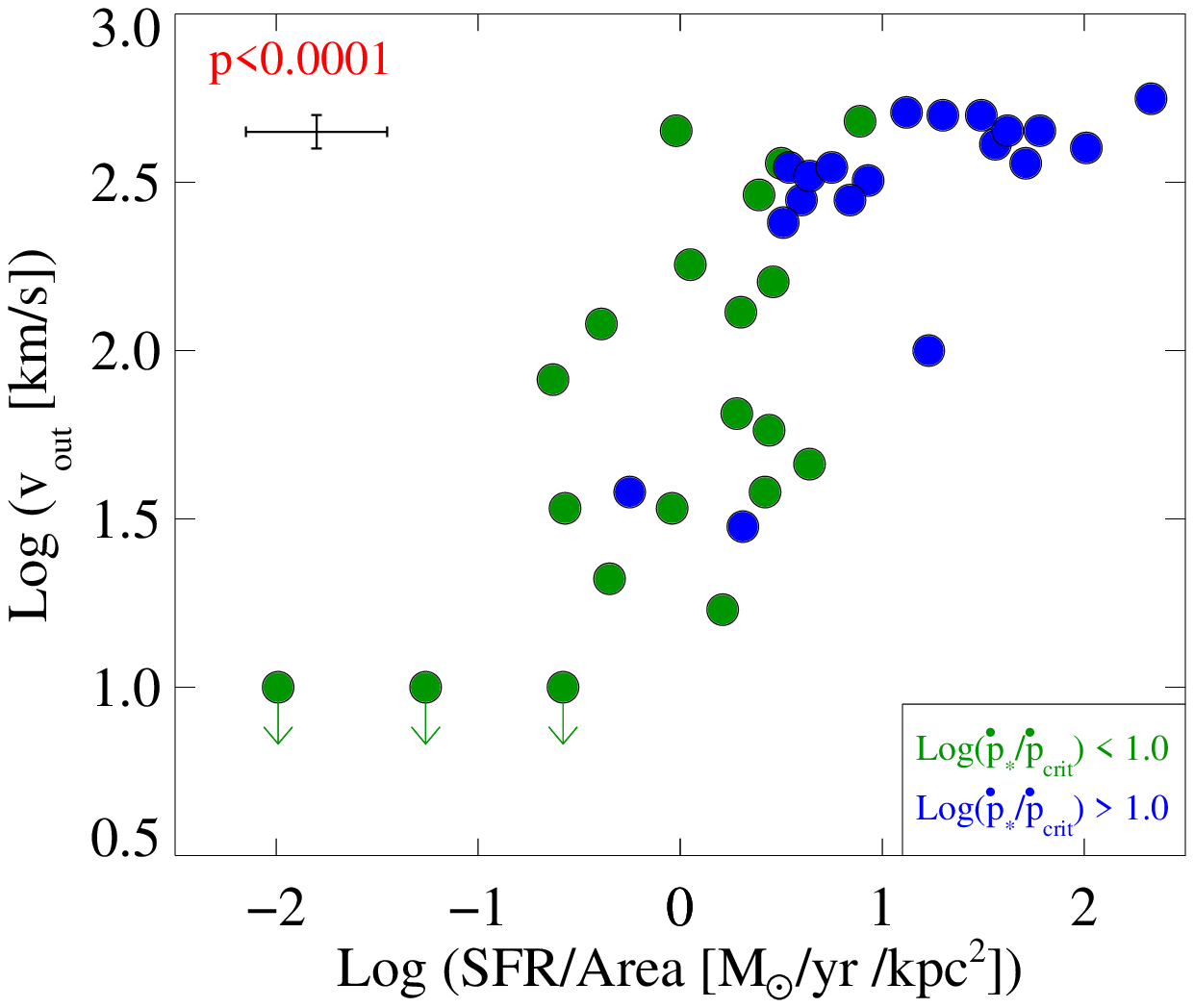}
\end{center}
\caption{The log of the outflow velocity is plotted as a function of the basic properties of the starburst galaxies.  Clockwise from the upper left (a,b,c,d). The upper left panel (a) shows that there is only a weak correlation between the outflow velocity and the galaxy circular velocity. The diagonal lines show $v_{out} = 10~ v_{cir }$, $v_{out} = v_{cir}$, $v_{out} = 0.1 v_{cir}$. The ratio $v_{out}/v_{cir}$ spans a range of over two orders-of-magnitude. The label on the upper axis shows the corresponding values of the galaxy stellar mass (see text). The upper right panel (b) shows a significant correlation between SFR and $v_{out}$, as has also been seen in other previous studies (see text).  The bottom two panels show the correlation with two forms of normalized SFR: SFR/area (c) and SFR/M$_*$ (d). Both correlations are significant, but the correlation with SFR/area is stronger. The crosses represent the typical uncertainties (see Table notes). The blue and green points show the strong- and weak-outflows. See section 4.1 and Fig. 9a.}  
 \end{figure*}

\subsubsection{Outflow Velocities}
It is usually thought to be straightforward to measure outflow velocities using the observed absorption-line profiles. However, as emphasized by Prochaska et al. (2011) and Scarlata \& Panagia (2015), the UV absorption-lines are resonance lines and as such, each upward transition resulting in absorption should be followed by a downward radiative transition resulting in emission. This emission can infill the absorption-line, and can in principle have a significant effect on the net observed profile. Both of the above papers show that the fine-structure transitions that are associated with many of the commonly-used resonance lines can be used to deduce the properties of the infilling resonance line emission. These effects have been investigated observationally in several recent papers (Martin et al. 2012; Erb et al. 2012; Kornei et al. 2012; Rubin et al. 2014). 

A similar approach to these observational papers has been adopted by A15, who have explicitly shown that the effect of infilling has a negligible effect on the properties of the outflows inferred from the observed absorption-line profiles in the COS sample. We have inspected the individual spectra of the FUSE sample, looking for the N II$^*$ 1085.7 and N III$^*$ 991.5 fine structure emission-lines (associated with the N II 1084.0 and N III 989.8 resonance lines). No significant detections were found, and therefore infilling will not be significant in these galaxies either.

Given that the ionized gas is the dominant phase in the outflows (see above), we have chosen to characterize the outflow velocity using the strongest relevant lines available.  For the COS data, this is the Si III 1206 line (A15). In Figure 1 we show line profiles for several strong interstellar absorption lines taken from the high signal-to-noise, smoothed ($R \sim 4000$) stacked spectrum of the COS sample described in A15 (N II 1084, Si III 1206, Si II 1260, C II 1334, Si IV 1393). It is clear that the Si III 1206 line is representative of the profiles of the other strong lines. For the FUSE data, the strongest line tracing ionized gas is CIII 977, but the line measured with the highest S/N ratio is NII 1084 (G09). We take the un-weighted mean of the measured outflow speeds using these two lines. As seen in Figure 1, the N II absorption-line in the stacked COS spectrum has the same properties as the Si III line, so there should be no systematic differences between the outflow velocities measured in the FUSE {\it vs.} COS samples.

In all cases we use a non-parametric measure of the outflow velocity based on the flux-weighted line centroid defined relative to the systemic velocity of the galaxy (G09; A15).  We emphasize that, absent a good model for the radial dependence of the mass-weighted distribution of the outflow velocity field, any specific definition of a characteristic outflow velocity is somewhat arbitrary. While a different outflow velocity from ours could be adopted (for example, the 'maximum' value where the line meets continuum, or the velocity enclosing the 80\% red-most portion of the integrated line profile), such parameters are usually more difficult to measure (more subject to uncertainties in the continuum placement and more dependent on the signal-to-noise in the data). Our choice is based on ease and robustness of measurement, and does not imply some specific model for the outflow. In any case, using a different definition would not affect the qualitative sense of our results.

The line centroids are significantly blue-shifted in all 21 of the LBAs with COS data and in 15 of the 18 FUSE starbursts.  This underscores the ubiquity of outflows in local starbursts. These outflow velocities are given in Table 2.

\begin{figure*}
\begin{center}
\includegraphics[trim=0mm 0mm 33mm 0mm,  clip=true,scale=0.55]{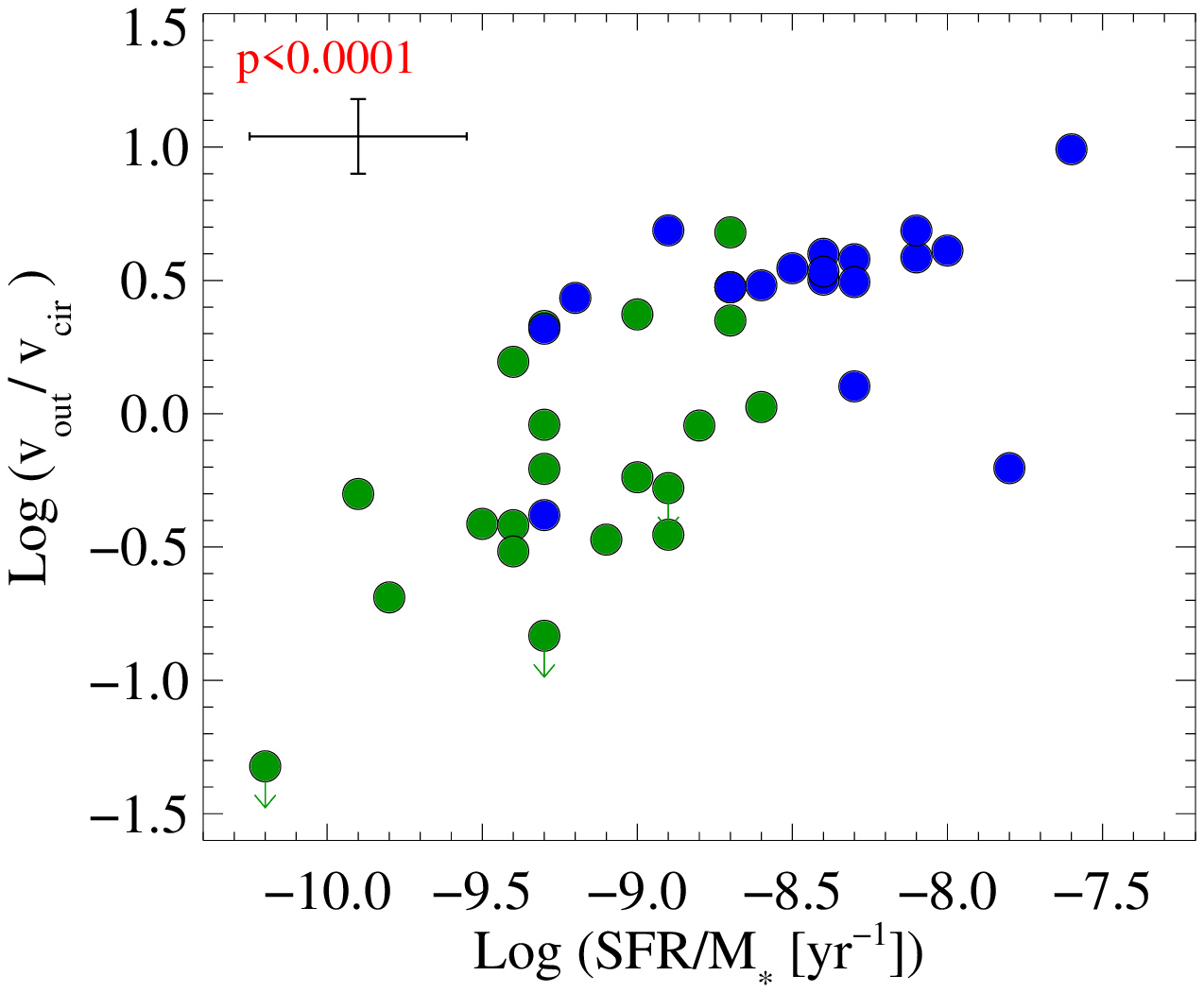} 
\includegraphics[trim=0mm 0mm 33mm 0mm,  clip=true,scale=0.55]{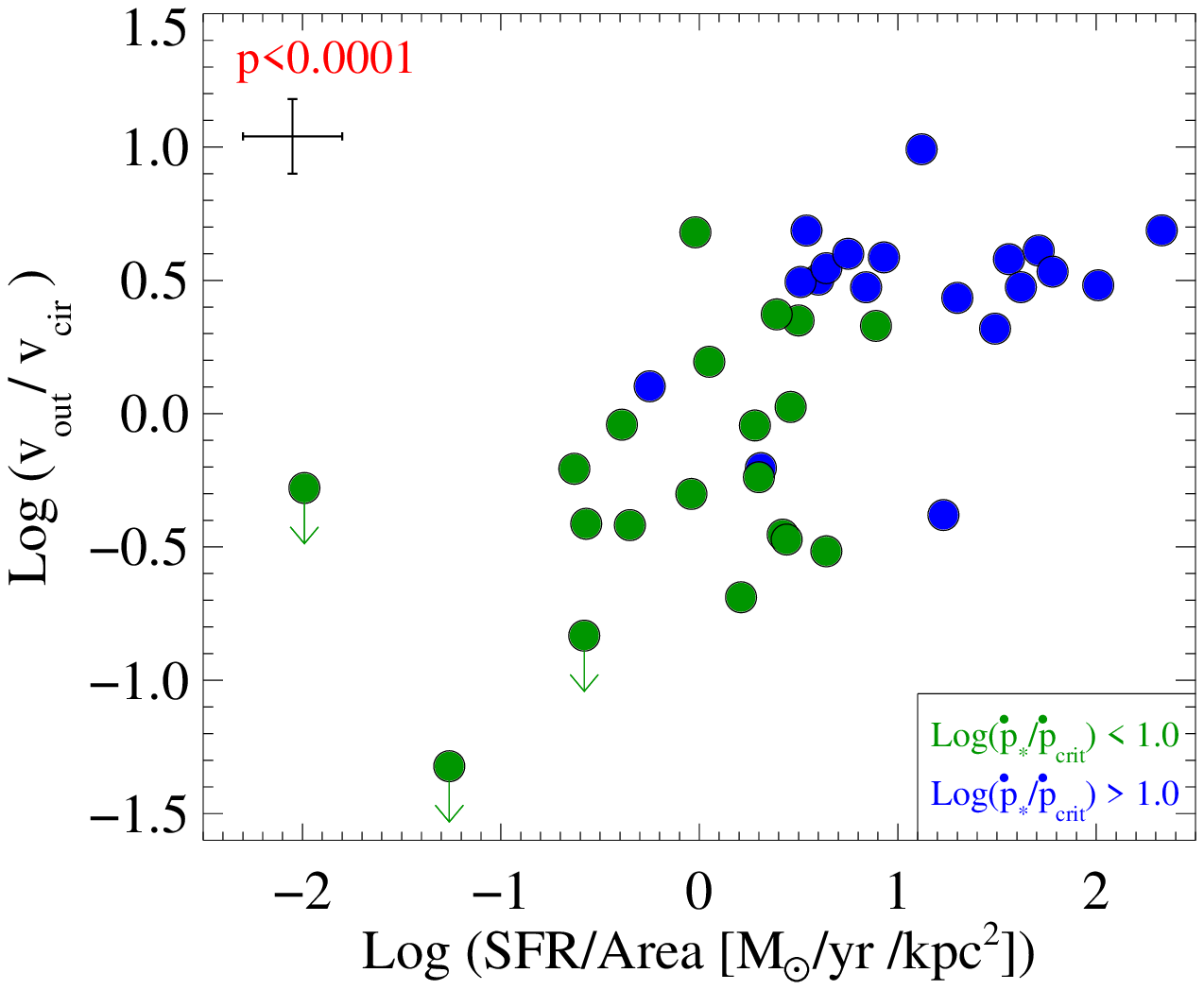}  
\end{center}
\caption{Left:  a) The outflow velocity normalized by the galaxy circular velocity plotted as a function of SFR/area. Right: b)  $v_{out}/v_{cir}$  plotted as a function of the specific SFR (SFR/M$_*$). Both plots show strong correlations. The crosses represent the typical uncertainties (see Table notes). The blue and green points show the strong- and weak-outflows. See section 4.1 and Fig. 9a.}  
 \end{figure*}

\begin{figure}
\begin{center}
\includegraphics[trim=0mm 20mm 0mm 10mm, scale=0.41]{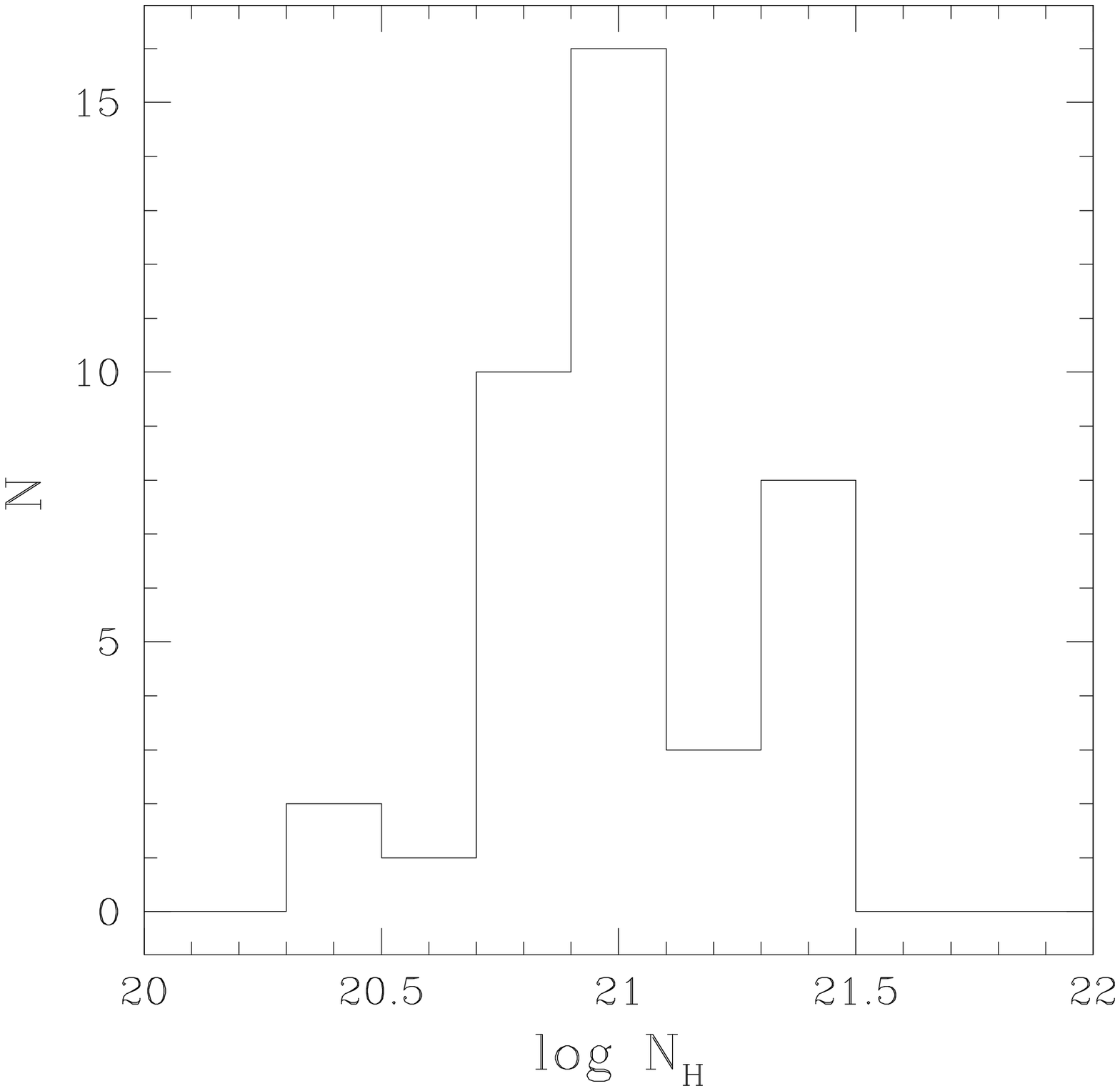}
\end{center}
\caption{A histogram of the total Hydrogen column density for the merged COS plus FUSE sample, estimated using the amount of the effective UV dust extinction,  the Calzetti et al. (2000) starburst attenuation law, and an assumption of a fixed (Milky Way) ratio of dust to metals in the ISM.  The typical column density of 10$\rm ^{21}~ cm^{-2}$ agrees with estimates based on the optically-thin interstellar absorption lines in these galaxies. See text for details.}
\end{figure}

\section{Empirical Correlations}
\subsection{Outflow Velocity}

We begin by examining the correlation of the outflow velocity (v$_{out}$) with the principal properties of the galaxy and its starburst.  Throughout the remainder of the paper we will use the Kendall $\tau$ test to assess the statistical significance of each correlation.  As shown in Figure 2, we see only a weak correlation with M$_*$  (and hence with $v_{cir}$).  There is a statistically significant positive correlation with SFR, SFR/M$_*$ , and SFR/area.  The last correlation is the strongest, and the impression is that the outflow velocity climbs rapidly with SFR/area but then saturates at about 400 km s$^{-1}$ for SFR/area $>$ 6 M$_{\odot}$ yr$^{-1}$ kpc$^{-2}$. We will return to this result in section 4 below where we compare the data to predictions of idealized wind models.

The correlation between outflow velocity and star-formation rate (albeit with significant scatter) has been previously seen in samples of low-redshift starbursts (Heckman et al. 2000; Martin 2005; Rupke et al. 2005; Martin et al. 2012), with a form that is qualitatively consistent with our Figure 2b. Studies at higher-redshift have been more mixed, but this most likely reflects the rather small range in star-formation rate that has been probed (Weiner et al. 2009; Erb et al. 2012; Kornei et al. 2012; Rubin et al. 2014; Bordoloi et al. 2014). There has also been disagreement about whether the outflow velocity correlates with SFR/area (Chen et al. 2010; Kornei et al. 2012; Rubin et al. 2014), but again these studies have covered relatively small ranges in this parameter compared to our sample. 

The possible dependence of outflow velocity on either M$_*$ or v$_{cir}$ has been examined by many of these studies. Most have found either no correlation, or only a weak one (Heckman et al. 2000; Rupke et al. 2005; Erb et al. 2012; Martin et al. 2012; Rubin et al. 2014; Bordoloi et al. 2014). The strongest correlation was found by Martin (2005). Ours is the only investigation in which the dynamic range in these quantities was similar to hers. Unfortunately, her sample only consisted of nine galaxies. The galaxies with low stellar masses and circular velocities in her sample did not contain the population of objects with relatively high outflow velocities ($v_{out} \sim 10^{2.5}~\rm km~ sec^{-1}$) and low circular velocities ($v_{cir} < 10^2$ km sec$^{-1}$) seen in our Figure 2b. We emphasize that we only poorly sample the population of galaxies  with M$_* < \rm 10^8~ M_{\odot}$ ($v_{cir} < 35 km s^{-1}$) and cannot say whether the trends we see extend into this range.      

We can also consider the normalized outflow velocity (v$_{out}/v_{cir}$) . As seen in Figure 3, this ratio spans over two-orders-of-magnitude ($\sim <0.1$ to 10). There are significant positive correlations between the ratio and normalized measures of the star formation rate (SFR/area and SFR/M$_*$). The correlation with SFR/area also shows a suggestion of a possible saturation in normalized outflow velocity at $v_{out} \sim$ 2 to 4 $v_{cir}$ above SFR/area $\sim$ 10 M$_{\odot}$ yr$^{-1}$ kpc$^{-2}$ . It is important to emphasize that outflow velocities this high are well in excess of the galaxy escape velocities. Moreover, these velocities refer to the centroid of the absorption-line, while outflowing gas is observed at velocities at least twice this high (e.g. Figure 1).

\subsection{Outflow Rates}
In order to turn the outflow velocities into outflow rates, several additional pieces of information are needed. More specifically, the mass outflow rate is given by $\dot{M} =  \Omega N_{out}  <m> v_{out} r_{out}$. Here, $\Omega$ is the solid angle occupied by the outflow,  N$_{out}$ is the total column density of the outflow along the line-of-sight,  $<m>$ is the mean mass per particle, and $r_{out}$ is the characteristic radius of the location of the absorbing material in the outflow. Given the ubiquity of outflows in our sample, we will set $\Omega = 4 \pi$ (while acknowledging that our UV-bright sample may be selectively missing galaxies viewed near the equatorial plane of a bi-polar outflow – see Heckman et al. 2000; Chen et al. 2010; Kornei et al. 2012; Rubin et al. 2014). In the sections below we will describe the estimates of the column densities and radii of the winds.

\subsubsection{Estimated Column Densities \& Wind Radii}
 In section 2.2.1 we used weak optically-thin absorption-lines to estimate column densities in the outflows. To check whether the total inferred column densities are plausible, we have used an indirect method that is independent of these estimates.  We first defined an effective far-UV opacity due to dust grains: $\tau_{dust} = - ln (L_{UV,obs}/L_{UV,int})$., where the UV luminosities refer to the observed (transmitted) and intrinsic (incident) quantities.   We define L$_{UV,int} = L_{UV,obs} + L_{IR,tot}$, where  L$_ {IR,tot} $ is the luminosity between wavelengths of 8 and 1000 $\micron$ and L$_{UV,obs} = 1.66 L_{FUV}$ is the luminosity shortward of 3000 \AA\ (Meurer, Heckman, \& Calzetti (1999). See A15 and G09 for a description of the relevant data. The resulting values for $\tau_{dust}$  are listed in Table 1. The starbursts have a mean value $\tau_{dust} = 1.5$ with a dispersion of 1.0 (these objects are translucent in the UV).

\begin{figure}
\begin{center}
\includegraphics[trim=0mm 10mm 0mm 10mm,scale=0.60]{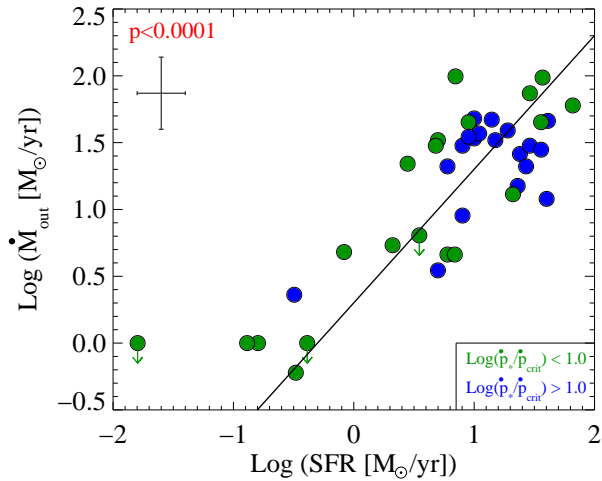}
\end{center}
\caption{The estimated outflow rate of ionized gas plotted as a function of the SFR (both in M$_{\odot}$ year$^{-1}$). The two quantities are well-correlated. The diagonal line shows a `mass-loading factor' ($\dot{M}_{out}$/SFR) of two, close to the median value for the sample. The cross represents the typical uncertainties (see Table notes). The blue and green points show the strong- and weak-outflows. See section 4.1 and Fig. 9a.}
\end{figure}

\begin{figure*}
\begin{center}
\includegraphics[trim=0mm 0mm 33mm 0mm,  clip=true,scale=0.59]{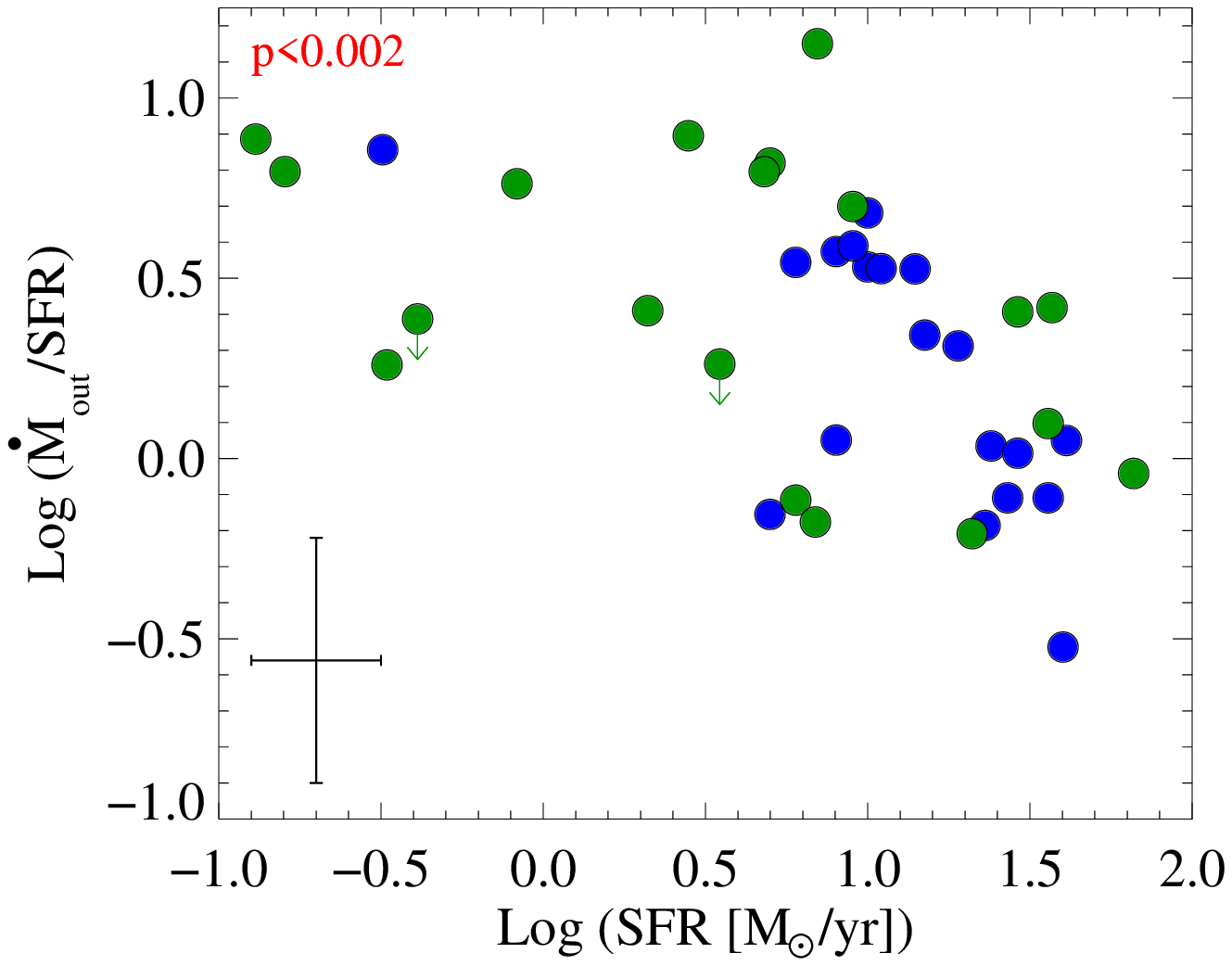}
\includegraphics[trim=0mm 0mm 33mm 0mm,  clip=true,scale=0.59]{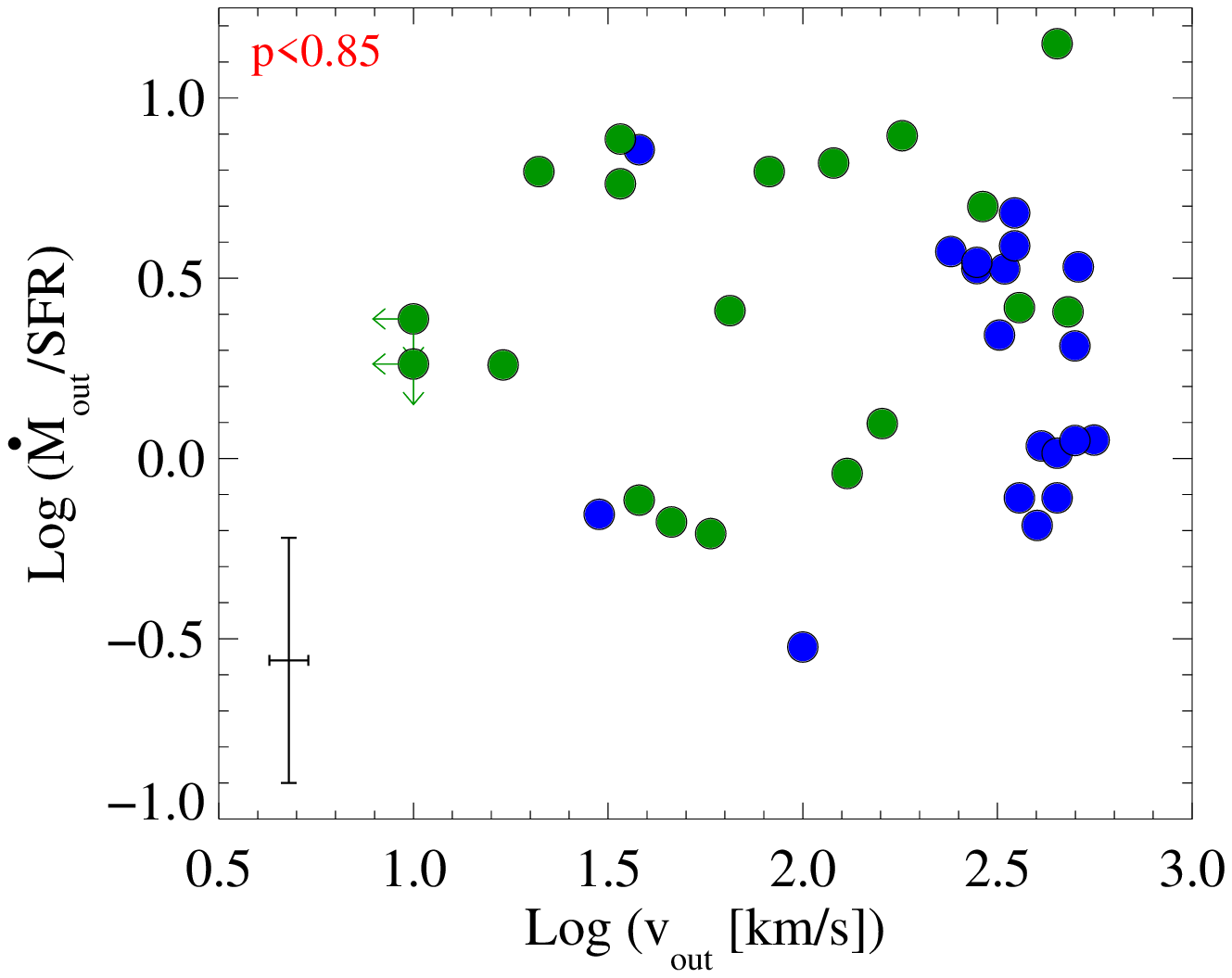}\\
\includegraphics[trim=0mm 0mm 33mm 0mm,  clip=true,scale=0.59]{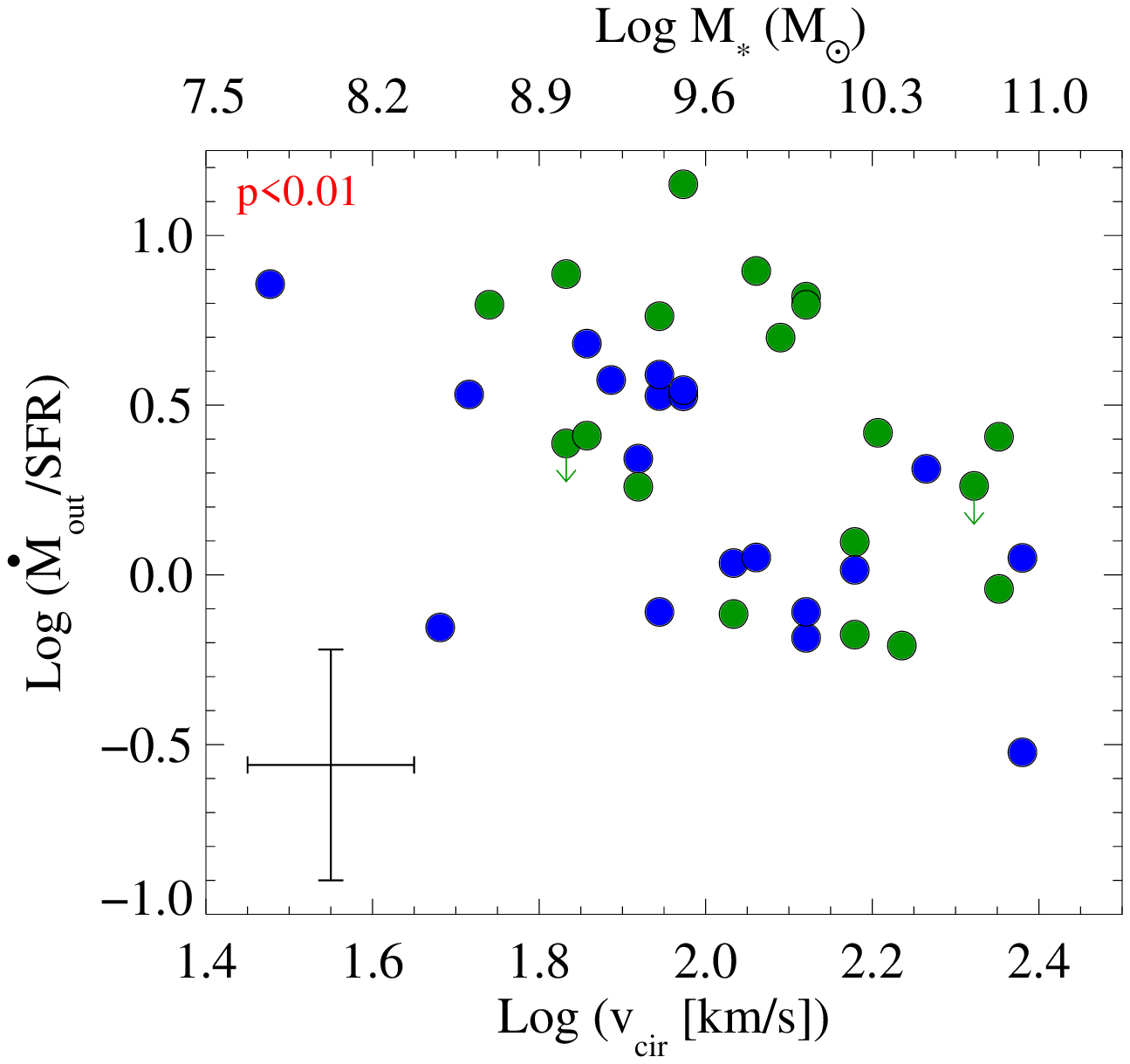}
\includegraphics[trim=0mm 0mm 33mm 0mm,  clip=true,scale=0.59]{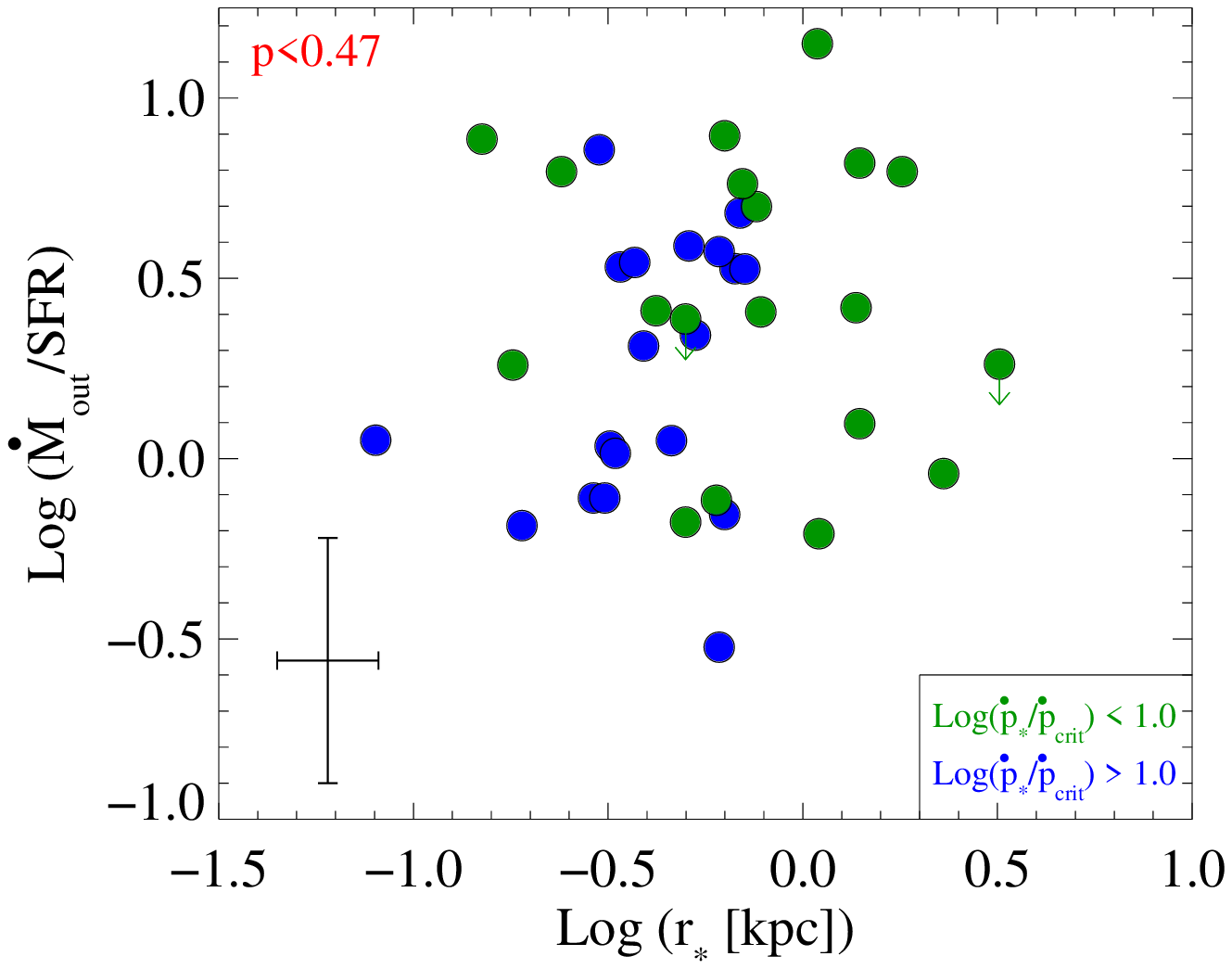}
\end{center}
\caption{The mass-loading factor ($\dot{M}_{out}/SFR$) is plotted as function of the basic properties of the galaxy and its starburst.  Clockwise from upper left (a,b,c,d). The two left panels (a,d) show weak inverse correlations of mass-loading with (top) SFR and (bottom) galaxy circular velocity and stellar mass. The two right panels (b,c) show that there is no correlation with either the outflow velocity (top) or the starburst radius (bottom). See text for details. The crosses represent the typical uncertainties (see Table notes). The blue and green points show the strong- and weak-outflows. See section 4.1 and Fig. 9a.}
\end{figure*}

The Calzetti starburst dust attenuation law (Calzetti et al. 2000) then yields the column density of dust corresponding to a given value of $\tau_{dust}$, and we convert this into a total Hydrogen column density assuming that all the galaxies have the same fraction of metals residing in dust grains (0.5) as in the Milky Way (see Mattsson et al. 2014 and references therein). The implied column densities are listed in Table 2 and shown in histogram form in Figure 4. The distribution has a mean of N$_H = 10^{21.0}$ cm$^{-2}$ with a standard deviation of only 0.25 dex. While this method is clearly a crude one, it is reassuring that this column density is similar to the values derived more directly using the weak absorption lines.  It also suggests that there is not a large range in total column density within our UV-selected sample of galaxies. Higher column densities (few $\times 10^{21}$ to $10^{22}$ cm$^{-2}$) have been inferred in the dusty outflows seen in Na I optical absorption-lines in far-IR-selected samples (Heckman et al. 2000; Rupke et al. 2005; Martin 2005).

The radius to be used to calculate the outflow rate depends upon the assumed model for the absorbing gas. In the idealized case of a thin shell-like structure, it is simply the radius of that shell. For a mass-conserving outflow with $v_{out}$  independent of radius, the appropriate value to use is that of the starting point of the wind. For the classic case of  an outflow of gas clouds driven by the ram pressure of a hot wind fluid created from thermalized supernova ejecta (Chevalier \& Clegg 1985), this would be the sonic radius of the wind (roughly, the radius of the starburst).  Since the correct value of the radius is uncertain, we will simply assume that it is some scale factor $\beta$ times the starburst radius r$_*$.

\begin{figure}
\begin{center}
\includegraphics[trim=0mm 3mm 33mm 10mm, clip=true, scale=0.60]{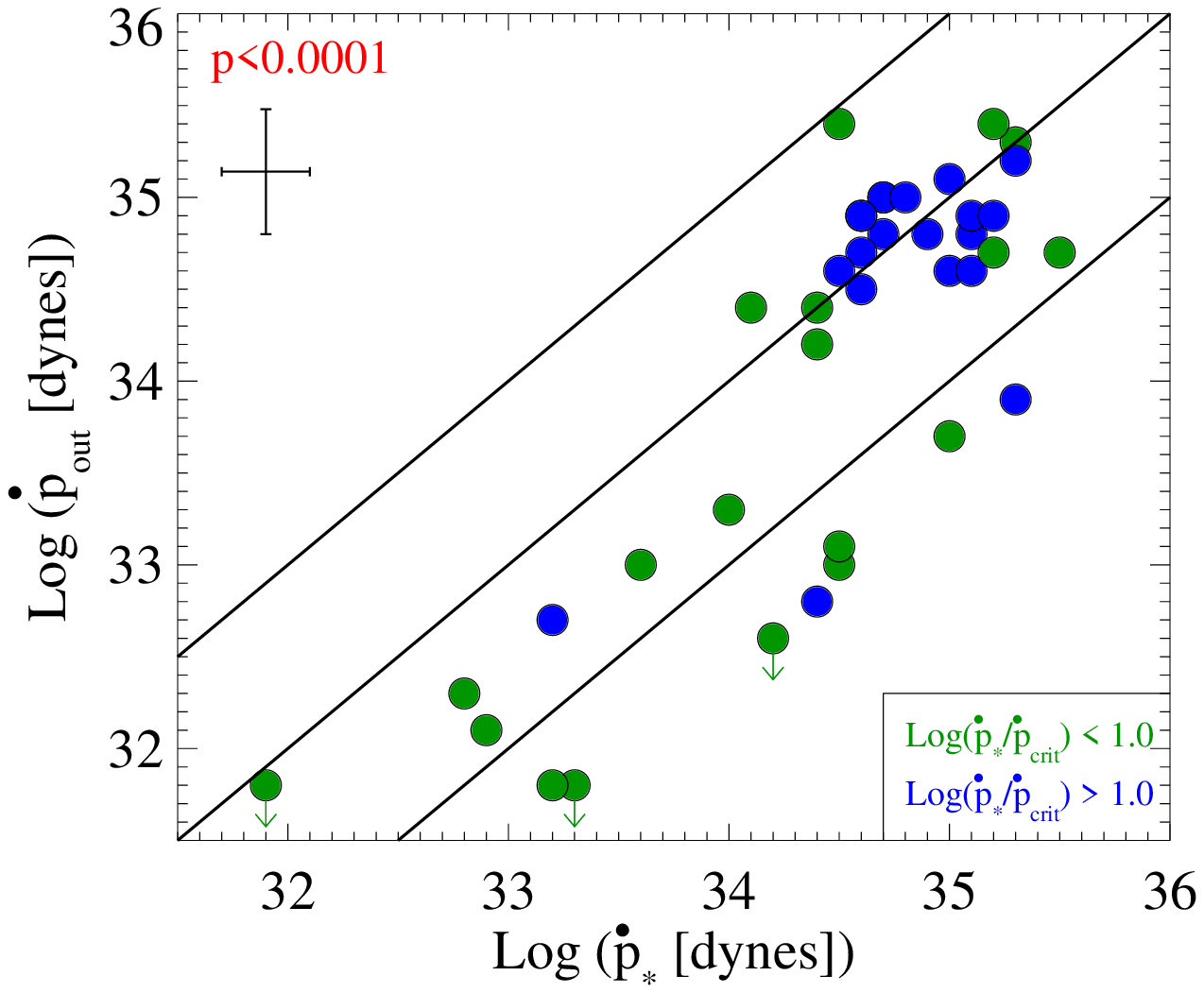}
\end{center}
\caption{The estimated observed rate of momentum transport in the outflow is plotted as a function of the rate of momentum supplied by the starburst (in dynes). The diagonal lines show the ratios $\dot{p}_{out}/\dot{p}_*$ = 10, 1, and 0.1. The forceful outflows ($\dot{p}_{out} > 10^{34}$ dynes) roughly carry the full amount of the momentum supplied by the starburst. The less forceful outflows carry typically only $\sim 10\%$ of the available momentum. The cross represents the typical uncertainties (see Table notes). The blue and green points show the strong- and weak-outflows. See section 4.1 and Fig. 9a.}
\end{figure}

\subsubsection{Resulting Correlations}
Given the above, it is clear that the calculated outflow rates are uncertain, and should only be taken as roughly indicative estimates that are prone to systematic uncertainties. With that caveat, we list the estimated outflow rates in Table 2 and show the relationship between  $\dot{M}_{out}$ and the SFR in Figure 5.  There is a significant correlation. The plot shown assumes a total wind column H density of 10$\rm ^{20.85}~ cm^{-2}$ (the mean of the estimates given above) and $\beta = 2$.  For these choices, the outflow rates typically range between 1 and 4 times the SFR (see Table 2). Similar results have been found previously for dusty low-z starbursts (e.g. Heckman et al. 2000; Rupke et al. 2005; Martin 2005), for intermediate redshift UV-selected galaxies (Bordoloi et al. 2014), and for higher-redshift Lyman Break Galaxies (Pettini et al. 2002).

The ratio of $\dot{M}_{out}$/SFR is referred to as the mass-loading factor. In Figure 6 we show that the mass-loading factor shows no significant dependence on $v_{out}$ or r$*$, and only weak (inverse) dependences on M$_*$, or $v_{cir}$.  The apparent inverse correlation between $\dot{M}_{out}/SFR$ with SFR must be treated with caution, since this a plot of $y/x$ {\it vs.} $x$ (which can induce a non-physical correlation). We will discuss the significance of these results in section 4 below.

We can also estimate the momentum flux carried by the outflow:  $\dot{p}_{out} = \dot{M} v_{out}$. These values are listed in Table 2.  We can now compare these values to the total momentum flux supplied by the starburst. This would be due to a combination of a hot wind fluid driven by the thermalized ejecta of massive stars (Chevalier \& Clegg 1985) and radiation pressure (Murray et al. 2005).  Based on Starburst99 (Leitherer et al. 1999), for a typical starburst population, the ratio of the momentum flux from these two respective sources is about 2.5. Summing them together yields a total momentum flux $\dot{p}_* = 4.8 \times 10^{33}$ SFR dynes, where the SFR is in M$_{\odot}$ year$^{-1}$.  The resulting values are given in Table 2. In Figure 7 we find a strong but non-linear correlation between $\dot{p}_{out}$ and $\dot{p}_*$ (see also Rupke et al. 2005).  We will discuss the significance of this below.

\section{Discussion}
\subsection{Comparisons to Simple Analytic Wind Models}
In this section we will compare the observed outflow velocities with predictions of several idealized analytic models for the outflow.

\subsubsection{Momentum-Driven Population of Clouds}
One simple model (e.g. Chevalier \& Clegg 1985) is that the absorption-lines are produced by a population of clouds/filaments that are driven outward by the momentum supplied by the starburst, as quantified above. In this picture, each cloud would be smaller than the starburst, and the observed absorption-line profile would reflect the distribution in radial velocity of the cloud population located along the lines-of-sight to the starburst. Such a picture is consistent with optical imaging of the ionized {\it emission-line} gas seen in nearby starburst-driven outflows (e.g. Mutchler et al. 2007) and with high-resolution numerical simulations of such flows (e.g. Cooper et al. 2008, 2009). A model of this kind would require enough clouds along the lines-of-sight to the starburst to produce the smooth absorption-line profiles that are observed in our galaxies (A15, G09).

The outward force on a cloud with a cross-sectional area $A_{c}$ located a distance $r$ from the starburst is:
\begin{equation}
F_{out} = A_{c} \dot{p}_*/4 \pi  r^2
\end{equation}
In addition to the outward force, a cloud with mass $M_c$ will also experience an inward force due to gravity:	
\begin{equation}
F_{in} = M_{c}  v_{cir}^2  r^{-1}
\end{equation}
where $v_{cir} = {G M(<r) /r}^{1/2}$ and $M(<r)$ is the total mass interior to $r$. Writing $M_{c}$ as $A_{c} N_{c}  <m>$ (where $N_c$ is the cloud Hydrogen column density and $<m>$ is the mean mass per H particle), we can then define a critical momentum flux such that the net force on a cloud located at $r = r_*$ is outward:
\begin{equation}
	\dot{p}_{crit,c} = 4 \pi r_* N_{c}  <m> v_{cir}^2
\end{equation}
In convenient units, $\dot{p}_{crit,c}$ is $10^{33.9}$ dynes for $N_{c} = \rm 10^{21}~cm^{-2}, ~r_*= 1$ kpc, and $v_{cir} = \rm 100~ km~sec^{-1}$.  The resulting values for $\dot{p}_{crit,c}$ are listed in Table 2. In Figure 8 we plot this critical momentum flux {\it vs.} the momentum flux available from the starburst. We then see that in 95\% of the cases (37/39), the starburst momentum flux exceeds the critical value for an outflow (often by over an order-of-magnitude).   

Taking this a step further, Figure 9 plots the normalized outflow velocity v$_{out}/v_{cir}$  {\it vs.}  $\dot{p}_*/\dot{p}_{crit,c}$ (hereafter $R_{crit,c}$).  This shows a strong positive correlation.  A closer examination suggests that $v_{out}/v_{cir}$ at first increases with increasing $R_{crit,c}$ , but then saturates at about  v$_{out}/v_{cir} \sim$ 3 to 5 for  ratios of $R_{crit,c} > 10$.  Crudely speaking, this would suggest three regimes: a `no-outflow' regime where $\dot{p}_* < \dot{p}_{crit,c}$, a `weak-outflow' regime where $\dot{p}_* \sim 1-10 ~\dot{p}_{crit,c}$, and a `strong-outflow' regime, where $\dot{p}_* > 10 \dot{p}_{crit,c}$. In this regime the outflow exceeds the galaxy escape velocity. To illustrate their different properties, we have color-coded the strong-outflows (blue) and weak/no-outflows (green) in Figures 2, 3, 5, 6, and 7. We will return to the implications of some of these differences in section 4.3 below.

We now compare the predicted and observed outflow velocities. To do this, we adopt a simple model in which a given cloud starts at some initial radius at rest (we take r$_*$) and is then accelerated by the combined force due to gravity and the starburst momentum flux. This yields the following expression for the velocity of the cloud as a function of its distance from the nucleus ($\beta r_*$):
\begin{equation}
v_{c}(\beta)/v_{cir} = \sqrt{2}[(1 - \beta^{-1})(R_{crit,c}) - ln(\beta)]^{1/2}
\end{equation}

The key importance of $R_{crit,c}$ is clear in this expression. In the limit $R_{crit,c} >>$ 1, gravity becomes negligible and for $\beta >> 1$ the expression for $v_c$ simplifies to that given in Chevalier \& Clegg (1985). One possible explanation for the observed spread of radial velocities of the clouds (as seen in the absorption-line profiles) would be a range in cloud column densities (since $\dot{p}_{crit,c} \propto N_c$). 

Because the momentum flux acting on a cloud drops like $1/r^2$, while the gravitational force in our assumed isothermal potential drops like $1/r$, the outbound cloud will eventually reach a maximum velocity at the radius where these two forces are equal (which will occur at $\beta = R_{crit,c}$). This allows us to compute a maximum outflow velocity for a given cloud:
\begin{equation}
v_{max,c}/v_{cir} = \sqrt{2} [(R_{crit,c}-1)-ln(R_{crit,c})]^{1/2}
\end{equation}

\begin{figure*}
\hspace{-.5cm}
\includegraphics[trim=0mm 0mm 33mm 0mm, clip=true, scale=0.61]{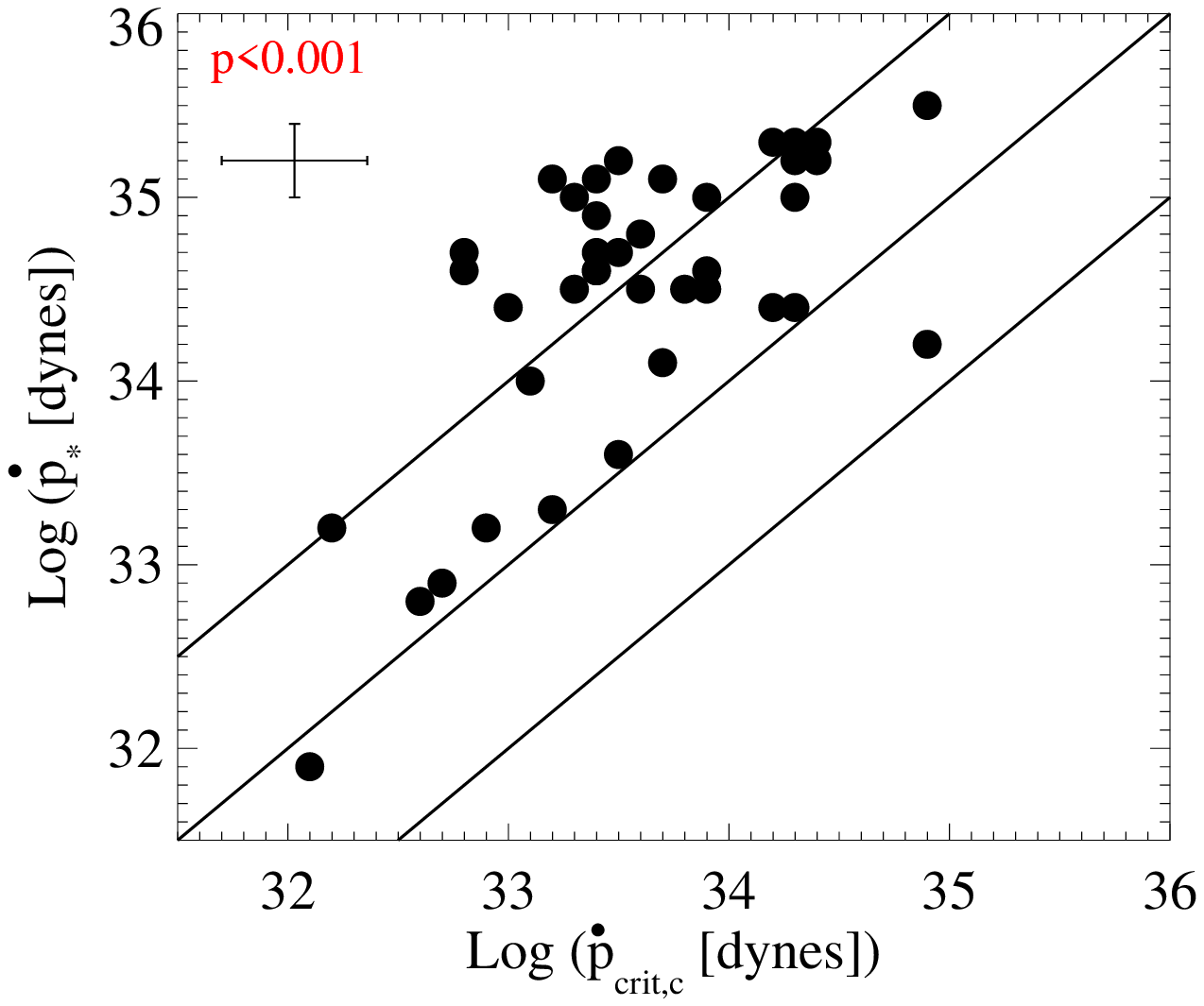}
\includegraphics[trim=0mm 0mm 33mm 0mm, clip=true, scale=0.61]{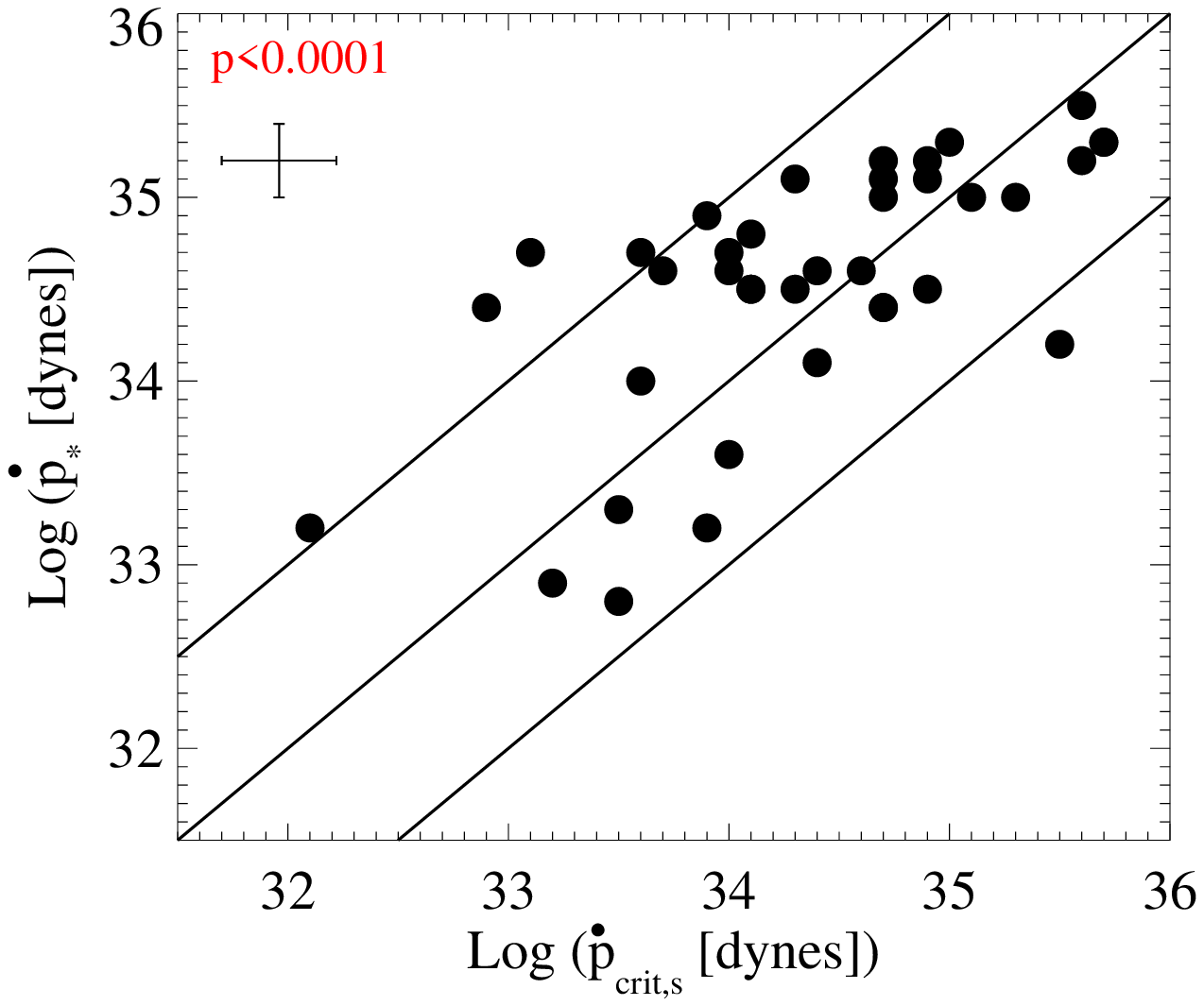}
\caption{The rate of momentum supplied by the starburst plotted as function of the critical value of momentum flux needed to overcome gravity, and to hence drive an outflow (see text). The left panel (a) applies to a model of a population of clouds and the right panel (b) to a spherical shell of gas. The diagonal lines show $\dot{p}_*/\dot{p}_{crit} =$ 10, 1, and 0.1. For the cloud model roughly 95\% of the galaxies lie above the critical value, compared to only 60\% for the shell model. Outflows are observed in about 90\% of the starbursts in our sample.
The crosses represent the typical uncertainties (see Table notes).}
\end{figure*}

We overplot this equation on the data shown in Figure 9 (using $v_{max,c}/2$ to represent the predicted centroid of absorption-line tracing the outflow). The predicted and observed velocities are in satisfactory agreement for most of the data points. This is gratifying, given the simplicity of the model. In particular, the model allows us to understand the steep rise in $v_{out}/v_{cir}$ for relatively small values of $R_{crit,c}$ (a regime where the $ln(R_{crit,c})$ term is significant), and a roll-over at large values where the model asymptotes to $v_{out}/v_{cir} = v_{max}/2v_{cir}= [R_{crit,c}/2]^{1/2}$.

We also note that Equation 4 predicts that clouds are rapidly accelerated. It is straightforward algebra to show that a cloud initially at rest at the starburst radius ($\beta = 1$) achieves half of $v_{max,c}$ defined by Equation 5 at a radius $\beta \leq 4/3$. This rapid radial rise in $v_{out}(\beta)$ is consistent with spatially-resolved maps of the outflows of ionized emission-line gas in nearby starbursts (e.g. Shopbell \& Bland-Hawthorn 1998).

\subsubsection{Momentum-Driven Shell}
This model is similar to the one above, but assumes that the absorption arises in an expanding spherical shell of gas centered on the starburst. Just as in the case of the cloud, there is an outward force produced by the momentum supplied by the starburst, here given simply as $F_{out} = \dot{p}_*$.  There is also an inward force due to gravity, given by $F_{in} = M_{s}  v_{cir}^2  r_{s}^{-1}$. We neglect pressure terms here. Defining $f_s$ to be the ratio of shell mass $M_s$ to the total mass interior to $r$, and then substituting for $M_s$ in the above equation we get $F_{in} = f_s v_{cir}^4/G$. This expression for $F_{in}$ then defines  the critical value for $\dot{p}_*$  for the net force to be outward.  In convenient units this is $\dot{p}_{crit,s} =  10^{34.2}$ dynes for $f_s = 0.1$ and v$_{cir} = 100$ km sec$^{-1}$. These values are listed in Table 2.

In Figure 8, we plot this quantity {\it vs.} $\dot{p}_*$ assuming that $f_s = 0.1$. This plot is rather similar to the one for the cloud model, but here only 59\% (23/39) of the galaxies lie above the critical value.  We then plot $v_{out}/v_{cir}$ {\it vs. }$\dot{p}_*/\dot{p}_{crit,s}$ in Figure 9.  While this plot is qualitatively similar to that for the cloud, the scatter here is significantly larger (and the correlation is not statistically significant).

Solving for the equation of motion for a shell we get:
\begin{equation}
v_s(\beta)/v_{cir}  = \sqrt{2}[(\dot{p}_*/\dot{p}_{crit,s} -1)  ln(\beta) ] ^{1/2}
\end{equation}
In this case, the ratio of outward to inward force is independent of radius, and the shell does not reach a maximum velocity.
In Figure 9 we overplot the predicted value of $v_s/v_{cir}$ {\it vs.} the observed ratio $v_{out}/v_{cir}$ assuming that $\beta =2$. The match to the data in this plot is significantly worse than for the cloud model. We conclude that this model is less successful than the cloud model in predicting both which galaxies will have outflows, and the observed outflow velocities.

\begin{figure*}
\hspace{-.5cm}
\includegraphics[trim=0mm 0mm 33mm 0mm,  clip=true,scale=0.6]{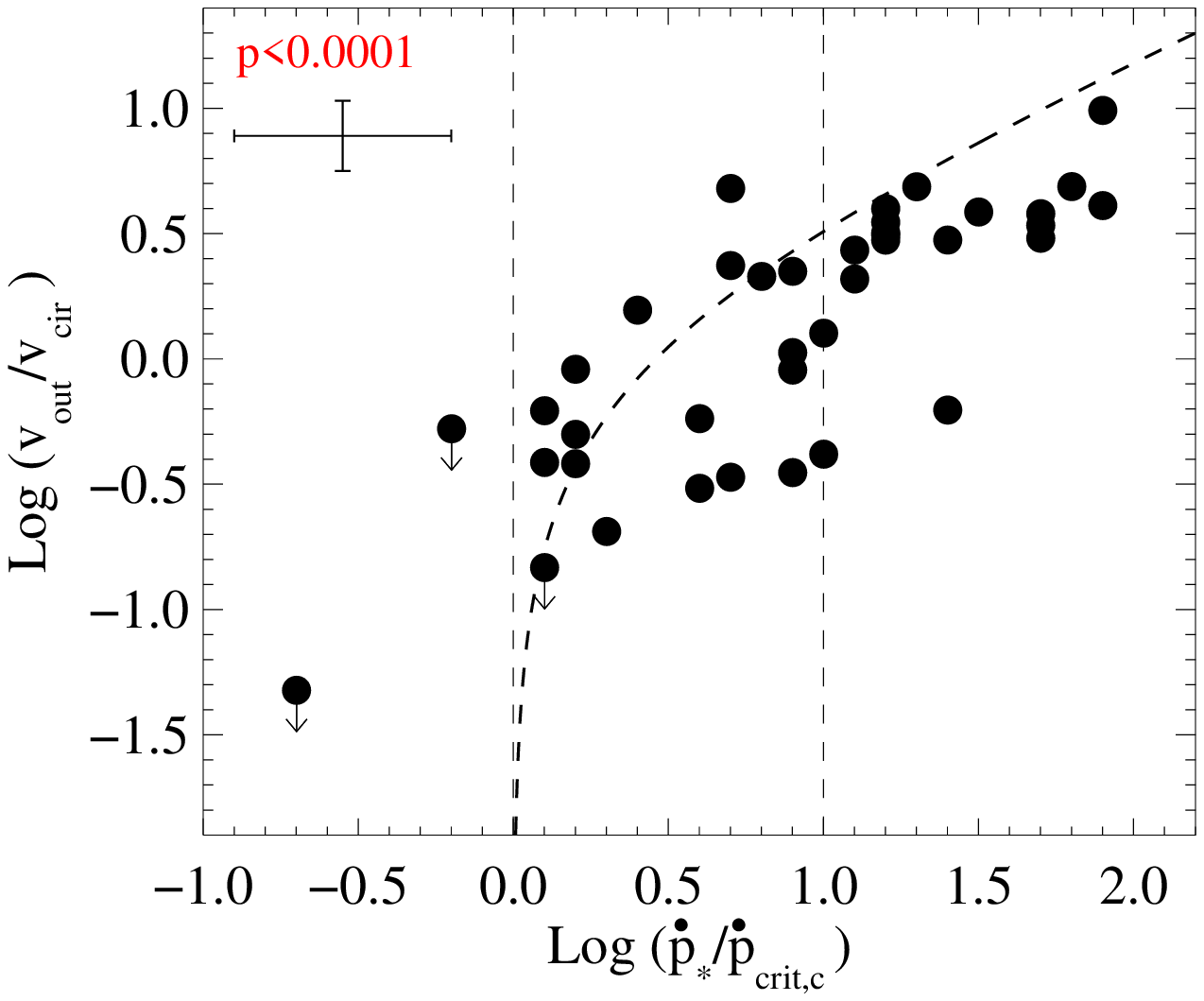}
\includegraphics[trim=0mm 0mm 33mm 0mm,  clip=true,scale=0.6]{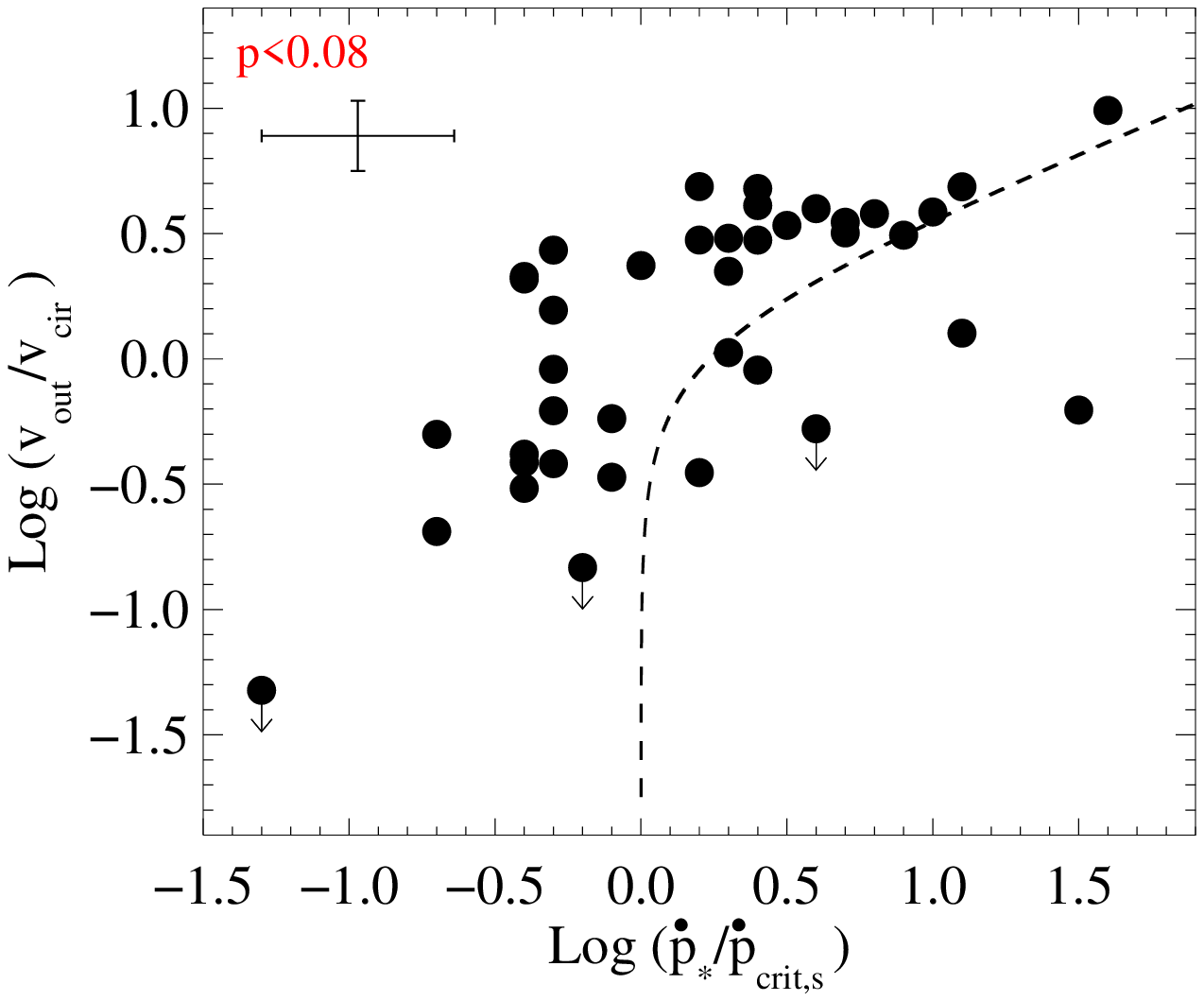}
\caption{The observed normalized outflow velocity ($v_{out}/v_{cir}$) is plotted as a function of the ratio of the amount of momentum flux supplied by the starburst to the critical value needed to overcome gravity and drive an outflow (see text). The left (a) and right (b) panels show the respective cases for the cloud and shell models.  The cloud model shows three regimes (indicated by dashed vertical lines). A `no-outflow' regime ($\dot{p}_* < \dot{p}_{crit}$), a `weak-outflow' regime in which $v_{out}/v_{cir}$ increases rapidly with increasing $\dot{p}_*/\dot{p}_{crit}$, and a 'strong-outflow' regime above $\dot{p}_*/\dot{p}_{crit} \sim 10$ where $v_{out}/v_{cir}$ is typically 3 to 5. The over-plotted curve is derived from the equation-of-motion for a momentum-driven cloud, and agrees with the data in most cases. The right panel for the shell model is qualitatively similar but shows much greater scatter and does a poorer job of predicting which starbursts drive outflows. The plotted curve is derived from the equation-of-motion for the shell model, and is generally a poor fit to the data. See text. The crosses represent the typical uncertainties (see Table notes).}
\end{figure*}

\subsection{Further Considerations}
We concluded above that the momentum-driven clouds model does a better job in reproducing the properties of observed outflows, and now consider further aspects of this model.  We found $R_{crit,c}$ (the ratio of the momentum flux supplied by the starburst to the momentum flux necessary to overcome gravity) is a key parameter. We have then divided the sample into strong-outflows ($R_{crit,c} >$ 10) and weak-outflows ($R_{crit,c} <$ 10).  

To further understand these results, we return to our result in section 3 where we compared the momentum flux in the observed outflow to that injected by the starburst.  In Figure 7 we found a strong but non-linear correlation between $\dot{p}_{out}$ and $\dot{p}_*$.  The most forceful outflows ($\dot{p}_{out} > 10^{34}$ dynes)  usually carry a significant fraction of the starburst output (of order 100\%).  The less forceful outflows carry a much smaller fraction (of order 10\%).  
These results can be understood in the following way.  For an outflow that carries all the momentum flux supplied by the starburst we have $\dot{M}_{out} v_{out} = 10^{33.7}$ SFR dynes (where SFR is in M$_{\odot}$ per year). Dividing both sides by $\dot{M}_{out}$ and converting SFR to cgs units, we get v$_{out} = \rm 760~SFR/\dot{M}_{out}~km~s^{-1}$. The most forceful outflows reach $v_{out} \sim$ 300 to 500 \kms, implying typical mass loading factors ($\dot{M}_{out}/$SFR) of  $\sim $2 (consistent with the range shown in Figures 5 and 6).

Figure 6 also shows that the mass-loading factor is either only weakly dependent on, or independent of, the bulk properties of the galaxy ($\rm M_*$, v$_{cir}$), the starburst (SFR, r$_*$), or the wind ($v_{out}$). This would then imply that that the weak-outflows carry roughly the same (normalized) mass flux as the strong-outflows, they just do so at a significantly lower velocity and therefore carry a relatively small fraction of the momentum flux supplied by the starburst. This presumably reflects the fact that inward gravitational forces are significant in these cases.  In Figure 10 we show that in nearly every case, the strong-outflows ($\dot{p}_* > 10 \dot{p}_{crit}$)  carry a significant fraction (of-order unity) of the momentum flux available from the starburst.  This is not generally true for the other starbursts.

Finally, we can re-examine the strong empirical correlations of the observed normalized outflow velocity ($v_{out}/v_{cir}$) with both and SFR/area and SFR/M$_*$ (Figure 2).   So long as the $ln(R_{crit,c})$ term in Equation 5 is insignificant (i.e. $R_{crit,c} >> 1$), the expression can be rewritten as:
\begin{equation}
(v_{max,c}/v_{cir})^2 \sim 2 R_{crit,c} = 2~\dot{p}_*/\dot{p}_{crit,c}
\end{equation}

Since $\dot{p}_{*} \propto SFR$, $\dot{p}_{crit,c} \propto r_* v_{cir}^2$, and $v_{cir} \propto M_{*}^{0.29}$,  equation 6 implies:
\begin{equation}
v_{max,c}/v_{cir}  \propto [(SFR/(r_* M_*^{0.58})]^{1/2}
\end{equation}

Given this expression, it is not surprising that the observed normalized outflow velocity correlates well with both $SFR/r_*^2 $ {\it and} $SFR/M_*$ (Figure 3).

\begin{figure}
\begin{center}
\includegraphics[trim=0mm 0mm 33mm 0mm,  clip=true,scale=0.61]{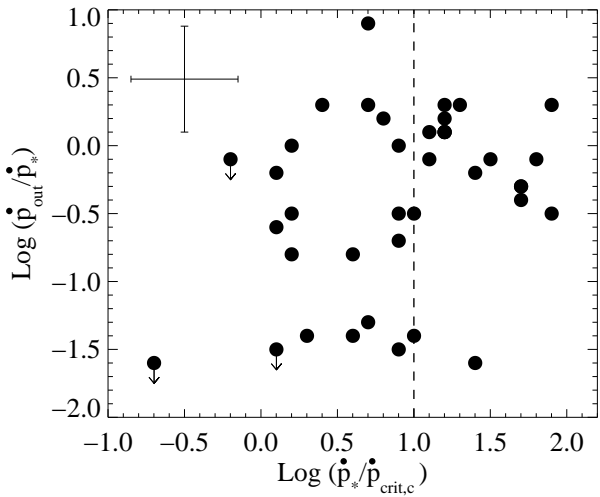}
\end{center}
\caption{The ratio of the momentum flux available from the starburst to the momentum flux in the observed outflow is plotted as function of the ratio of the starburst momentum flux to the critical value required to overcome gravity. This figure shows that most of the strong 'saturated' outflows ($\dot{p}_* > 10~ \dot{p}_{crit}$  - to the right of the dashed vertical line)  carry a significant fraction of the available momentum flux, while the others generally do not. See text. The crosses represent the typical uncertainties (see Table notes).}
\end{figure}

\subsection{Implications for `Sub-Grid Physics'}
As described in section 1, both cosmological numerical simulations and semi-analytic models for galaxy evolution often make use of simple prescriptions to deal with the effects of feedback from stars (`sub-grid physics'). The current state of such models have been reviewed by Somerville \& Dav{\'e} (2015). We close the paper by briefly summarizing the implications of our results for these models.

\begin{figure*}
\begin{center}
\includegraphics[trim=0mm 0mm 33mm 0mm,  clip=true,scale=0.60]{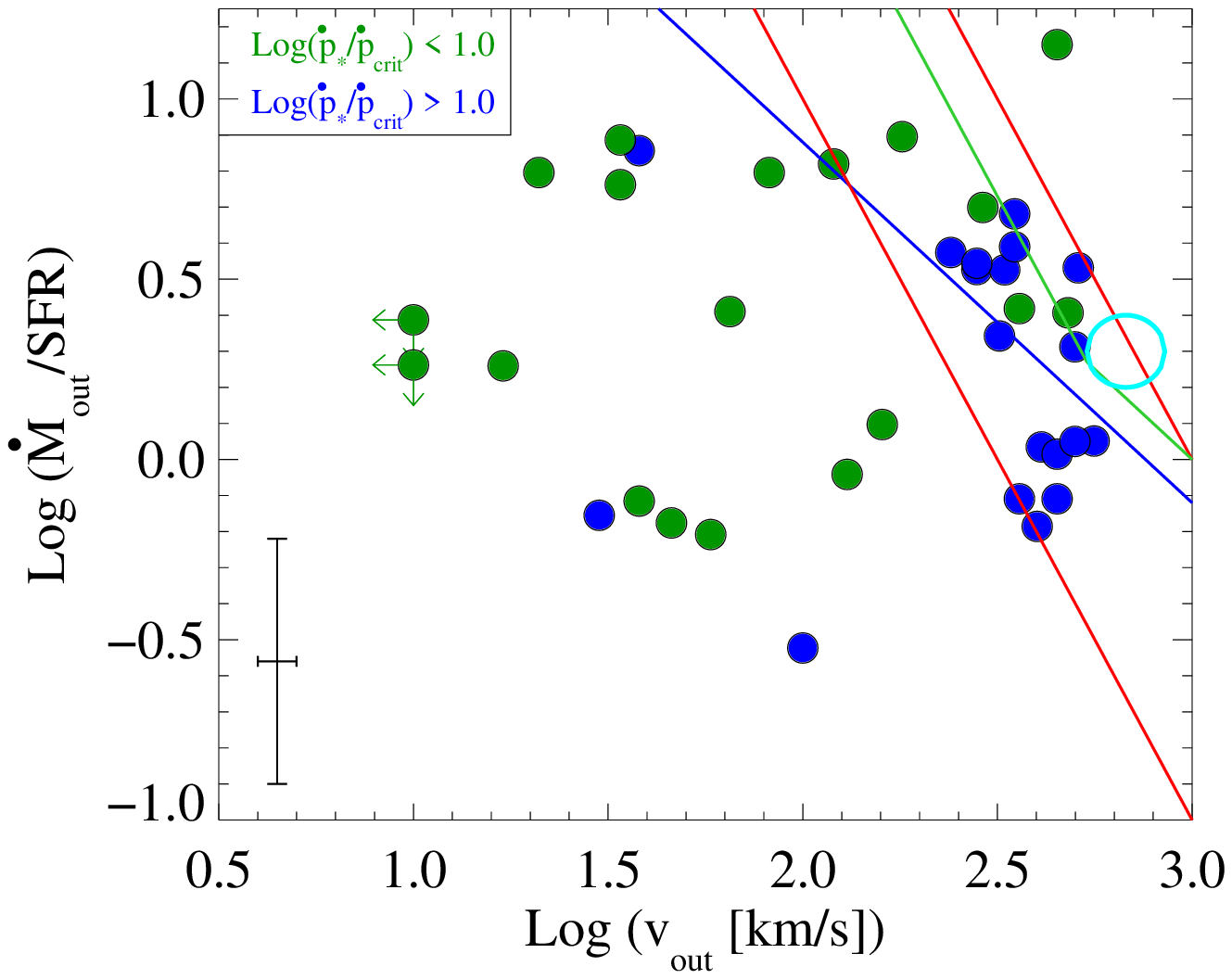}
\includegraphics[trim=0mm 0mm 33mm 0mm,  clip=true,scale=0.60]{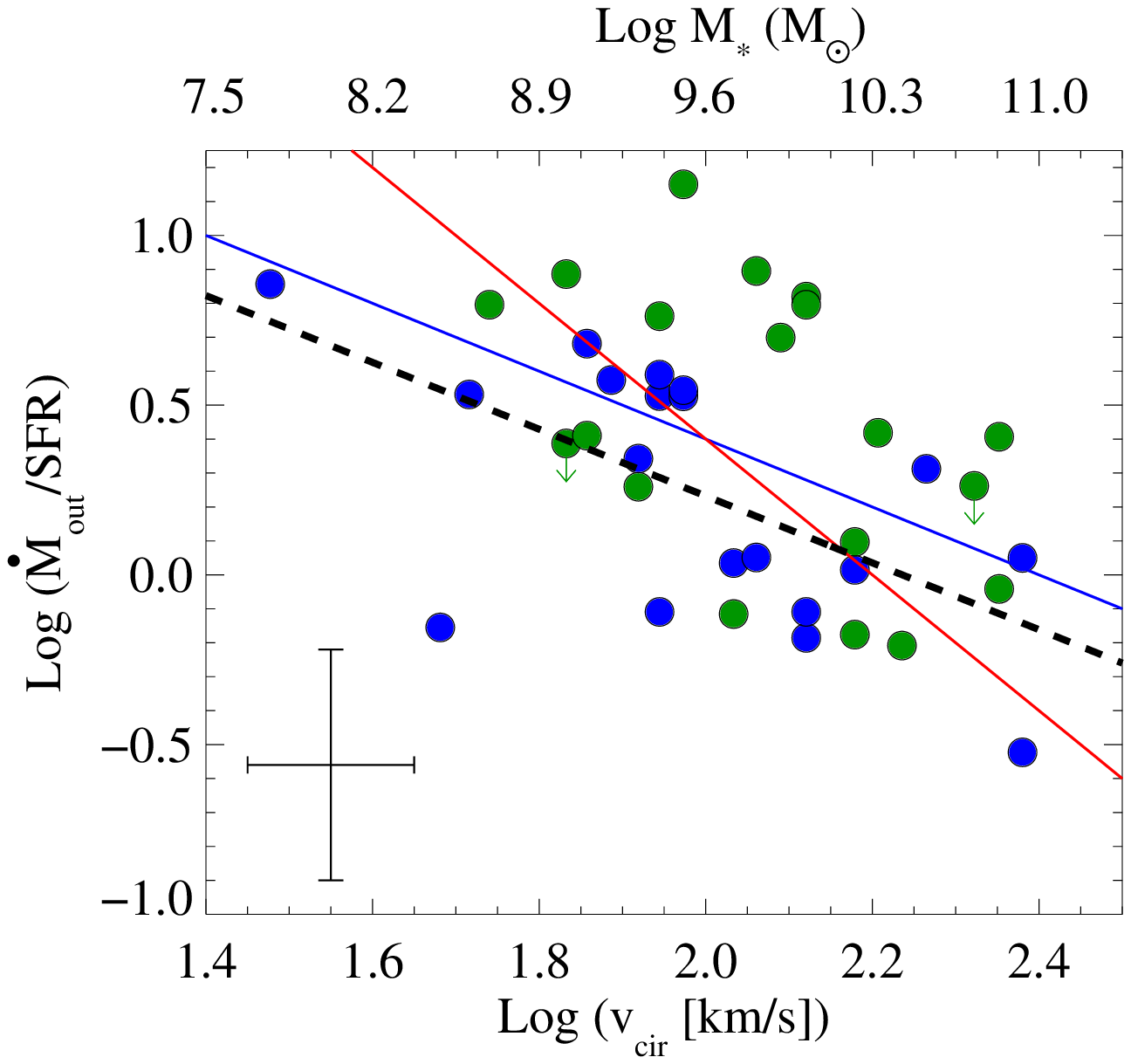}
\end{center}
\caption{Left (a): A magnified version of Figure 6b showing the observed relation between the mass-loading factor in the outflow ($\dot{M}_{out}/SFR$ {\it vs.} the outflow velocity $v_{out}$). The plotted lines refer to various assumptions made in semi-analytic models or cosmological simulations (see text). The blue line shows the assumption that the outflow contains 100\% of the momentum flux supplied by the starburst. The two red lines show the assumption that the outflow contains 100\% (upper) and 10\% (lower) of the kinetic energy flux supplied by the starburst. A hybrid `momentum-energy' model is shown as a broken green line. The large cyan hollow circle assumes a constant outflow velocity and mass-loading factor. The blue data points are for the strong-outflows ($R_{crit,c} >$ 10) and the green points are for the weak-outflows ($R_{crit,c} <$ 10). Only for the strong-outflows do the assumed relations roughly match the data. Right (b): A magnified version of Figure 6c showing the observed relation between mass-loading and the galaxy circular velocity $v_{cir}$. The blue and red diagonal lines schematically indicate scalings often made in models and simulations (with mass-loading proportional to $v_{cir}^{-1}$ and to $v_{cir}^{-2}$, respectively. The blue and green symbols indicate the strong- and weak-outflows respectively. The dashed black line shows a least-square fit to the strong-outflows.  See text for details. The crosses represent the typical uncertainties (see Table notes).}
\end{figure*}

\subsubsection{Comparing Data and Prescriptions}
We begin by comparing our data to predictions of outflow velocities.
One popular prescription is to assume that $v_{out} \propto v_{cir}$ (e.g. Baugh et al. 2005; Bower et al. 2006; Oppenheimer \& Dav{\'e} 2008; Dutton et al. 2010; Guo et al. 2011; Dav{\'e} et al. 2013). It is clear from Figures 1 and 2 that overall, this is a bad assumption: the ratio $v_{out}/v_{cir}$ spans about two orders-of-magnitude and has strong systematic dependences on both SFR/M$_*$ and SFR/area. Examination of Figure 9 shows that most of the scatter in the ratio of $v_{out}/v_{cir}$ is produced by the weak-outflows (R$_{crit,c} <$ 10). With a few exceptions, the strong-outflows (R$_{crit,c} >$ 10) have $v_{out} \sim 3 - 5 v_{cir}$. We have explained above how these results can be understood in the context of an outflow of a population of momentum-driven clouds.

Next, we consider our results on the relationship between mass-loading and outflow velocity. A common prescription is to assume that outflows use up the momentum flux supplied by the starburst, which results in a scaling such that SFR $\propto \dot{M}_{out} v_{out}$.  This is equivalent to assuming that the mass-loading factor in the outflow $\dot{M}_{out}$/SFR $\propto v_{out}^{-1}$ (e.g. Oppenheimer \& Dav{\'e} 2008; Dutton et al. 2010).  Figure 11a compares this scaling relation to the data, and shows it does not match the weak-outflows. As we have argued above, the scaling based on momentum conservation fails here because only the strong-outflows carry a substantial fraction of the available momentum flux. The weak-outflows are slow because they are ineffective at tapping this source of momentum, not because they are strongly mass-loaded. The strong-outflows are largely consistent with this prescription. 

A related prescription is to assume the outflows carry a fixed fraction of the kinetic energy supplied by the starburst, implying that the mass-loading factor scales like $v_{out}^{-2}$ (e.g. Baugh et al. 2005; Somerville et al. 2008).  Figure 11a shows that the strong-outflows of typically carry about 10 to 50\% of the available kinetic energy flux. This fraction can be much smaller in the weak-outflows in our sample. \footnote{We note that this does not necessarily imply that most of the kinetic energy input from the starburst has been lost through radiative cooling. The bulk of this energy could reside in the fast (but tenuous) wind fluid whose ram pressure is driving the outflow seen in the absorption-lines (see section 1).} The vast range in this fraction underscores the problem with adopting a simple proportionality between the input and output kinetic energy fluxes for all outflows.

Hybrid models with a mass-loading factor $\propto v_{out}^{-2}$ at low velocities and $\propto v_{out}^{-1}$ at higher velocities have also been adopted (Dav{\'e} et al. 2013; Vogelsberger et al. 2013; Ford et al. 2015). These have disagreements with the data that are similar to those of the simpler momentum-based and energy-based prescriptions (severely failing for the weak-outflows).

A different type of prescription assumes a fixed mass-loading factor ($\dot{M}_{out} \propto$ SFR) and that the outflows all have the same velocity (e.g. Springel \& Hernquist 2003; Schaye et al. 2010; Dav{\'e} et al. 2013; Ford et al. 2015).  As seen in Figure 11a, this prescription fails dramatically for the weak-outflows. We have shown that the outflow velocity depends on $\dot{p}_*/\dot{p}_{crit}$ in these cases.

As noted above, most prescriptions assume that the outflow velocity is proportional to the circular velocity. Thus, the proportionality between the mass-loading factor and the circular velocity in these models will be the same as that for the outflow velocity. This is shown more explicitly in Figure 11b. While we observe a general inverse correlation between the mass-loading term and circular velocity (as is assumed in most prescriptions), it important to emphasize that the weak-outflows are generally moving mass outward at velocities that are below the galaxy circular velocity (and certainly well below escape velocity). This is inconsistent with the assumptions in the model prescriptions. Considering only the strong-outflows (which are traveling at or above the escape velocity), we find a best-fit relation with a slope of -0.98 and unit mass-loading factor at $v_{cir} = 10^{2.25}$ km sec$^{-1}$. While this slope is consistent with a momentum-based scaling relation, the more direct test of such a relation is the plot of mass-loading {\it vs.} $v_{out}$ in Figure 11a. We also note that our empirical scaling relation is in good agreement (for $v_{cir} > 50$ km sec$^{-1}$)with the recent FIRE simulations by Muratov et al. (2015) evaluated at $z \sim$ 0. These simulations explicitly calculate stellar feedback rather than adopting a prescription.

In conclusion, some of the commonly adopted model prescriptions for outflow velocity and mass-loading agree with the observed properties of the strong-outflows. However, none of them adequately represent the properties of our full sample. In particular, the assumption that the outflow velocities simply scale with the galaxy circular velocity is incorrect for the weak-outflows, and the assumptions about how the mass-loading scales with the outflow velocity lead to disagreements with the data that are over an order-of-magnitude in many weak outflows.

\subsubsection{The Way Forward}
While current cosmological simulations with increasing numerical 
resolution are improving the capability to model outflows in a more realistic fashion (e.g. Cen 2014; Vogelsberger et al. 2013; Schaye et al. 2015; Henriques et al. 2015; Mitra et al. 2015), high-resolution simulations of individual galaxies that are able to calculate {\it ab initio} more of the relevant physical processes are clearly needed (e.g. Cooper et al. 2008, 2009; Shen et al. 2013; Hopkins et al. 2013; Muratov et al. 2015). Ideally, these simulations should be closely compared to (and influenced by) both the results in this paper and to detailed multi-phase observations of actual galactic winds in the local universe.

In the meantime, a formulation based on the results in this paper could be useful. That is (given a star-formation rate, a radius of the star-forming region, and a halo circular velocity) calculate $\dot{p}_*/\dot{p}_{crit,c}$ ($R_{crit,c}$). Use this to assign an outflow speed using our equation 5 above (and see Figure 9). It may be case that most stars over cosmic history form in systems with $R_{crit,c} > 10$ (leading to strong-outflows). In this case, the simple scaling in Equation 7 could be used: $v_{out} = v_{max,c}/2 \sim v_{cir} \sqrt{R_{crit,c}/2}$. 

The mass-loading term as defined in models/simulations is only really applicable to the strong-outflows, which are the only cases in which the outflow velocity typically exceeds the galaxy escape velocity. For these, the best-fit scaling relation for mass-loading shown in Figure 11b could be adopted. However, we emphasize that actual values of the mass-loading term were derived from our data under the assumption that the (absorption-weighted) size of the outflow is twice the radius of the starburst. A different scale factor will result in a different constant of proportionality between mass-loading and circular velocity for strong-outflows. 

\section{Conclusions}
We have analyzed outflows as traced by far-UV absorption lines for a combined sample of 39 low-redshift starburst galaxies. The majority of these starbursts (24) have properties that make them excellent local analogs to typical high-redshift Lyman Break Galaxies. This is in contrast with most previous investigations of low-z outflows, which have analyzed far-IR-selected starbursts using the optical Na I absorption-line. Compared to most studies of higher-redshift UV-selected galaxies, our individual spectra are higher in S/N and the sample spans much wider ranges in star-formation rates (SFR), galaxy stellar masses (M$_*$), circular velocities (v$_{cir}$), metallicities,  specific star-formation rates (SFR/M$_*$), and SFR/area.

We begin by summarizing the most robust empirical conclusions.  These all involve directly measured quantities without uncertain assumptions.  We found that the outflow velocity of the ionized gas seen in absorption correlates only weakly with the galaxy stellar mass (M$_*$) and the galaxy circular velocity ($v_{cir}$), but shows significant correlations with SFR/M$_*$, and (especially) with SFR and SFR/area.  Notably, the outflow velocity in our sample increases with increasing SFR/area, but appears to `saturate' at roughly 300 to 500 km s$^{-1}$ for SFR/area $\rm > 6 M_{\odot}\rm ~kpc^{-2}$.  The ratio of $v_{out}/v_{cir}$ spans nearly two orders-of-magnitude and correlates strongly and positively with both SFR/M$_*$ and SFR/area. This ratio reaches typical values of $\sim$ 3 to 5 for starbursts with high SFR/M$_*$ and SFR/area, well in excess of the galaxy escape velocity.

Our conclusions regarding outflow rates are more uncertain.  These require values for the total gas column density and characteristic radius of the outflow as seen in absorption.  We have used weak (optically-thin) metal absorption-lines to find that the column density of ionized gas is over an order-of-magnitude larger than of the neutral gas, and to show that the total implied hydrogen column density of-order 10$^{21}$ cm$^{-2}$ .  We simply parameterized the effective (absorption-weighted) radius of the absorbing gas as a multiple of the starburst radius ($r_{out} =  \beta r_*$). With a choice $\beta = 2$, the typical estimated mass outflow rates range from $\sim$ 1 to 4 times the SFR.  The `mass-loading' factor ($\dot{M}_{out}$/SFR) correlates only weakly and inversely with SFR, $v_{cir}$, or M$_*$, and shows no correlation with $v_{out}$ or r$_*$.

We then considered two simple analytic models for the outflow:  a population of clouds accelerated by the net sum of gravity and the momentum-flux supplied by the starburst, and a similar model in which the clouds were replaced by a spherical shell of gas.  We compared the momentum flux provided by the starburst ($\dot{p}_*$) to the critical value needed to overcome gravity ($\dot{p}_{crit}$) (which is different for the cloud and shell models). On this basis, the cloud model did a significantly better job of predicting both which galaxies should have outflows and the outflow velocities themselves.

Further consideration of the model of momentum-driven clouds led to some interesting insights. We found that the ratio ($R_{crit,c} = \dot{p}_*/\dot{p}_{crit,c}$) determines the observed outflow velocity. We identified three regimes: a `no-outflow' regime where $R_{crit,c} < 1$, a `weak-outflow' regime in which v$_{out}/v_{cir}$ increased with increasing $R_{crit,c}$, and then a `strong-outflow' regime ($R_{crit,c} > 10$) in which v$_{out}/v_{cir} \sim$ 3 to 5. We showed that this behavior can be rather well-represented by the equation-of-motion derived for this simple model of momentum-driven clouds. The strong-outflow regime corresponded to cases in which the estimated momentum flux in the outflowing gas ($\dot{p}_{out}$) is comparable to the total momentum flux supplied by the starburst ($\dot{p}_*$), and in which the outflows are faster than the galaxy escape velocity.  In the weak-outflow and no-outflow cases it is likely that gravity plays a significant role in the cloud dynamics and the coupling of the starburst momentum flux to outflowing gas is ineffective.

We showed that none of the simple prescriptions used to capture the essential features of winds in typical cosmological simulations and models of galaxy evolution agree with the empirical results for the full sample we have studied. In particular, there is no tight, simple scaling observed between $v_{out}$ and $v_{cir}$ (the ratio of the velocities spans nearly two orders-of-magnitude), and $v_{out}$ varies by over a factor of 50 (inconsistent with the so-called `constant wind' model). We see no evidence that the mass-loading factor ($\dot{M}_{out}$/SFR) correlates inversely with $v_{out}$, as is sometimes assumed. All these disagreements are severe for the weak-outflows. For most of the strong-outflows, at least some of the prescriptions can successfully describe the data within the uncertainties in our derived parameters.

We believe the empirical results in this paper can be understood on the basis of a simple physically-motivated model of a momentum-driven outflow of a population of clouds. Our results could be used to define a simple prescription for the ratio of outflow to circular velocity based on the value of $R_{crit,c}$ (which can be estimated from the model/simulated galaxy star-formation rate, the radius of the star-forming region, and the galaxy circular velocity). The mass-loading term can also be parameterized as being proportional to $v_{cir}^{-1}$, but only for the strong-outflows ($R_{crit,c} > 10$).

In closing, we emphasize that our hope is that the results presented in this paper will provide helpful guidance as to the treatment of galactic winds in future numerical simulations and theoretical models, as well as providing new insights into the nature of galactic winds driven by massive stars.

\vspace{.5cm}

\acknowledgements 
TH thanks the Kavli Institute for Theoretical Physics, the Simons Foundation, and the Aspen Center for Physics for supporting informal workshops on galactic winds, feedback, and the circum-galactic medium during 2014 and 2015. These provided much of the motivation for this paper.  Conversations at these workshops with Romeel Dav{\'e}, Crystal Martin, Norm Murray, Ralph Pudritz, Eliot Quataert, Brant Robertson, Chuck Steidel, Todd Thompson, and Ellen Zweibel were especially helpful. He also thanks the Sackler Foundation, Churchill College, and the Institute of Astronomy for supporting his visit to the University of Cambridge. He thanks Martin Haehnelt, Max Pettini, and Mark Whittle at the IoA for especially useful conversations. Finally, we thank the referee for suggestions that helped us clarify the presentation and discussion of our results.

This work is based on observations with the NASA/ESA Hubble Space Telescope, which is operated by the Association of Universities for Research in Astronomy, Inc., under NASA contract NAS5-26555. The data analysis was supported by grant HST GO 12603. This work was supported in part by National Science Foundation Grant No. PHYS-1066293 and the hospitality of the Aspen Center for Physics. This project also made use of SDSS data. Funding for the SDSS and SDSS-II has been provided by the Alfred P. Sloan Foundation, the Participating Institutions, the National Science Foundation, the U.S. Department of Energy, the National Aeronautics and Space Administration, the Japanese Monbukagakusho, the Max Planck Society, and the Higher Education Funding Council for England.  The SDSS Web Site is http://www.sdss.org/. 
The SDSS is managed by the Astrophysical Research Consortium for the Participating Institutions. The Participating Institutions are the American Museum of Natural History, Astrophysical Institute Potsdam, University of Basel, University of Cambridge, Case Western Reserve University, University of Chicago, Drexel University, Fermilab, the Institute for Advanced Study, the Japan Participation Group, Johns Hopkins University, the Joint Institute for Nuclear Astrophysics, the Kavli Institute for Particle Astrophysics and Cosmology, the Korean Scientist Group, the Chinese Academy of Sciences (LAMOST), Los Alamos National Laboratory, the Max-Planck-Institute for Astronomy (MPIA), the Max-Planck-Institute for Astrophysics (MPA), New Mexico State University, Ohio State University, University of Pittsburgh, University of Portsmouth, Princeton University, the United States Naval Observatory, and the University of Washington. 

{\it Facilities:}  \facility{Sloan ()} \facility{COS ()}

\clearpage
\LongTables 
\begin{landscape}
\begin{deluxetable}{ccccccccccccc}
\tabletypesize{\scriptsize}
\tablecaption{Description of Galaxy Properties. \label{tbl-galaxy}}
\tablewidth{0pt}
\tablehead{
\colhead{Galaxy} & \colhead{$\rm z$} & \colhead{\rm $r_*^a$} &\colhead{$\rm Log~ M_*^b$} & \colhead{$\rm  v_{cir}^c $} & \colhead{$\rm  SFR^d$} & \colhead{$\rm Log~ SFR/M_*$} & \colhead{$\rm Log~ SFR/area$}  & \colhead{$\rm 12 +log[O/H]^e$} & \colhead{$\rm \tau_{UV}^f$} & \colhead{Sample}\\
\colhead{}   & \colhead{}  & \colhead{(kpc)}  & \colhead{($\rm Log~M_{\odot}$)} &\colhead{($\rm km~s^{-1}$)}  & \colhead{($\rm M_ {\odot} yr^{-1}$) }  &\colhead{($\rm log~yr^{-1}$)}  & \colhead{($\rm M_{\odot} yr^{-1} kpc^{-2}$)} & \colhead{}& \colhead{}  & \colhead{}}
\startdata

J002101+005248 & 0.09840 & 0.53 & 9.3 & 83 & 15 & -8.1 & 0.93 & 8.20 & 0.8 & COS\\
J005527-002148 & 0.16744 & 0.32 & 9.7 & 108 & 24 & -8.3 & 1.56 & 8.28 & 2.1 & COS\\
J015028+130858 & 0.14668 & 1.37 & 10.3 & 161 & 37 & -8.7 & 0.50 & 8.39 & 2.8 & COS\\
J021348+125951 & 0.21902 &0.39 & 10.5 & 184 & 19 & -9.2 & 1.30 & 8.74 & 1.9 & COS\\
J080844+394852 & 0.09123 & 0.08 & 9.8 & 115 & 8 & -8.9 & 2.33 & 8.74 & 1.2 & COS\\
J082354+280621 & 0.04722 & 0.34 & 8.6 & 52 & 10 & -7.6 & 1.12 & 8.23 & 2.4 & COS\\
J092159+450912 & 0.23499 & 0.78 & 10.8 & 225 & 29 & -9.3 & 0.89 & 8.67 & 1.9 & COS\\
J092600+442736 & 0.18072 & 0.69 & 9.1 & 72 & 10 & -8.1 & 0.54 & 8.09 & 0.4 & COS\\
J093813+542825 & 0.10208 & 0.67 & 9.4 & 88 & 11 & -8.4 & 0.60 & 8.19 & 0.8 & COS\\
J102548+362258 & 0.12650 & 0.61 & 9.2 & 77 & 8 & -8.3 & 0.51 & 8.11 &  0.4 & COS\\
J111244+550347 & 0.13163 & 0.33 & 10.2 & 151 & 29 & -8.7 & 1.62 & 8.52 & 1.8 & COS\\
J111323+293039 & 0.17514 & 1.09 & 9.6 & 94 & 7 & -8.7 & -0.02 & 8.35 & 1.1 & COS\\
J114422+401221 & 0.12695 & 0.76 & 9.9 & 123 & 9  & -9.0  & 0.39 & 8.40 & 1.1 & COS\\
J141454+054047 & 0.08190 & 0.63 & 8.5 & 48 & 5 & -7.8 & 0.31 & 8.28 & 1.5 & COS\\
J141612+122340 & 0.12316 & 0.19 & 10.0 & 132 &  23 & -8.6 & 2.01 & 8.47 & 1.4 & COS\\
J142856+165339 & 0.18167 & 0.71 & 9.6 & 94 & 14 & - 8.5 & 0.64 & 8.31 & 0.7 & COS\\
J142947+064334 & 0.17350 & 0.29 & 9.4 & 88 & 27 & -8.0 & 1.71 & 8.12 & 1.4 & COS\\
J152141+075921 & 0.09426 & 0.37 & 9.5 & 94 & 6 & -8.7 & 0.84 & 8.27 & 0.25 & COS\\
J152521+075720 & 0.07579 & 0.51 & 9.4 & 88 & 9 & -8.4 & 0.75 & 8.46 & 1.5 & COS\\
J161245+081701 & 0.14914 & 0.31 & 10.0 & 132 & 36 & -8.4 & 1.78 & 8.51 & 1.7 & COS\\
J210358-072802 & 0.13689 & 0.46 & 10.9 & 240 & 41 & -9.3 & 1.49  & 8.70 & 2.7 & COS\\
Haro 11 & 0.02060 & 1.40 & 10.2 & 151 & 36 & -8.6 & 0.46 & 8.29 & 2.1 & FUSE (LBA)\\
VV 114  & 0.02007 & 2.30 & 10.8 & 225 & 66 & -9.0 & 0.30 & 8.65 & 2.6 & FUSE(LBA)\\
NGC 1140 & 0.00501 & 0.70 &  9.4 & 88 & 0.83 & -9.5 &  -0.57 & 8.27 & 0.7 & FUSE\\
SBS 0335-052 & 0.01349 & 0.30 & 7.8 & 30 & 0.32 & -8.3 & -0.25 & 7.25 & 0.1 & FUSE\\
Tol 0440-381 & 0.04100 & 1.4 & 10.0 & 132 & 5.0 & -9.3 & -0.39 & 8.20 & 0.6 & FUSE\\
NGC 1705 & 0.00211 & 0.24 & 8.6 & 55 & 0.16 & -9.4 & -0.35 & 8.07 & 0.15 & FUSE\\
NGC 1741 & 0.01347 & 0.60 & 9.7 & 108 & 6.0 & -8.9 & 0.42 & 8.27 & 1.1 & FUSE\\
I Zw 18 &  0.00251 & 0.50 &  7.1 & 19 & 0.016 & -8.9 & -1.99 & 7.20 & 0.1 & FUSE\\
NGC 3310 & 0.00328 & 0.63 &  9.8 & 115 & 2.8 & -9.4 & 0.05 & 8.40 & 1.8 & FUSE\\
Haro 3 & 0.00321 & 0.50 &  8.9 & 68 & 0.41 & -9.3 & -0.58 & 8.46 & 1.7 & FUSE\\
NGC 3690 & 0.01041 & 0.61 & 10.9 & 240 & 40 & -9.3 & 1.23 & 8.60 & 4.2 & FUSE\\
NGC 4214 & 0.00097 & 0.15 & 9.0 & 68 & 0.13 & -9.9 & -0.04 & 8.30 & 1.3 & FUSE\\
Mrk 54 & 0.04486 & 1.10 & 10.4 & 172 & 21 & -9.1 & 0.44 & 8.40 & 0.9 & FUSE(LBA)\\
M 83 & 0.00172 & 3.20 & 10.7 & 210 & 3.5 & -10.2 & -1.26 & 8.85 & 2.6 & FUSE\\
NGC 5253 & 0.00135 & 0.18 &  9.3 & 83 & 0.33 & -9.8 & 0.21 & 8.15 & 1.2 & FUSE\\
IRAS 19245-4140 & 0.00945 & 0.42 & 9.1 & 72 & 2.1 & -8.8 & 0.29 & 8.22 & 0.45 & FUSE\\
NGC 7673 & 0.01131 & 1.80 & 10.0 & 132 & 4.8 & -9.3 & -0.63 & 8.45 & 1.4 & FUSE\\
NGC 7714 & 0.00935 & 0.50 & 10.2 & 151 & 6.9 & -9.4  & 0.64 & 8.47 & 3.6 & FUSE

\enddata
\tablenotetext{a}{The half-light radius of the starburst as measured with UV imaging (see text for details). The measurement uncertainties are typically $<$ 0.1 dex.}
\tablenotetext{b}{The total galaxy stellar mass derived from near-IR and/or multiband optical photometry. Based on comparisons to the measured dynamical masses (Overzier et al. 2011), the typical uncertainties are $\pm$ 0.3 dex.}
\tablenotetext{c}{The characteristic circular velocity based on the fit between $M_*$ and $v_{cir}$ in Simons et al. (2015).  The scatter in this relationship is $\pm$ 0.1 dex.}
\tablenotetext{d}{The star-formation rate based on UV and IR photometry (see A15). A standard Kroupa/Chabrier IMF is assumed. Comparisons to SFRs derived from the extinction-corrected H$\alpha$ luminosity imply an uncertainty of $\pm$ 0.2 dex.}
\tablenotetext{e}{The oxygen abundance ($12+log[O/H]$) derived from the Pettini \& Pagel (2004) calibration of the O3N2 strong-line method. These authors quote an uncertainty of $\pm$0.14 dex.}
\tablenotetext{f}{The effective optical depth in the FUV defined as $-ln (L_{UV,obs}/L_{UV,int}$  (see text). The uncertainties are caused by systematic errors in $L_{UV,int}$ as derived from UV plus IR photometry. We estimate the resulting uncertainty on $\tau_{UV}$ to be $\sim \pm$ 0.2}
\end{deluxetable}
\clearpage
\end{landscape}

\clearpage
\LongTables 
\begin{landscape}
\begin{deluxetable}{ccccccccc}
\tabletypesize{\scriptsize}
\tablecaption{Description of Outflow Properties. \label{tbl-outflow}}
\tablewidth{0pt}
\tablehead{
\colhead{Galaxy}  & \colhead{\rm $v_{out}^a$} & \colhead{$\rm  \dot{M}_{out}^b$} & \colhead{$\rm  Log~ \dot{M}_{out}/SFR$} & \colhead{$\rm Log~ \dot{p}_{out}^c$} & \colhead{$\rm Log~ \dot{p}_*^d$} & \colhead {$\rm Log ~\dot{p}_{crit,c}^e$} & \colhead {$\rm Log~ \dot{p}_{crit,s}^f$} & \colhead{$\rm Log ~N_H$} \\
\colhead{}   & \colhead{($\rm km s^{-1}$)}   & \colhead{($\rm M_{\odot} yr^{-1}$)} &\colhead{}  & \colhead{($\rm log~ dynes$)} & \colhead{($\rm log~ dynes$)} &\colhead{($\rm log~ dynes$)} &\colhead{($\rm log ~dynes$)} & \colhead {($\rm log ~cm^{-2}$})}
\startdata
J002101+005248 & 320 & 33 & 0.35 & 34.8 & 34.9 & 33.4 & 33.9 & 21.0\\
J005527-002148 & 410 & 26 & 0.04 & 34.8 & 35.1 & 33.4 & 34.3 & 21.3\\
J015028+130858 & 360 & 97 & 0.41 & 35.3 & 35.3 & 34.4 & 35.0 & 21.3\\
J021348+125951 & 500 & 39 & 0.31 & 35.1 & 35.0 & 33.9 & 35.3 & 20.8\\
J080844+394852 & 560 & 9.0 & 0.02 & 34.5 & 34.6 & 32.8 & 34.4 & 20.6\\
J082354+280621 & 510 & 34 & 0.55 & 35.0 & 34.7 & 32.8 & 33.1 & 21.4 \\
J092159+450912 & 480 & 74 & 0.40 & 35.4 & 35.2 & 34.4 & 35.6 & 20.9 \\
J092600+442736 & 350 & 48 & 0.66 & 35.0 & 34.7 & 33.4 & 33.6 & 20.8 \\
J093813+542825 & 280 & 37 & 0.52 & 34.8 & 34.7 & 33.5 & 34.0 & 21.0 \\
J102548+362258 & 240 & 30 & 0.60 & 34.7 & 34.6 & 33.4 & 33.7 & 20.7 \\
J111244+550347 & 450 & 30 & 0.02 & 34.9 & 35.1 & 33.7 & 34.9 & 21.0 \\
J111323+293039 & 450 & 99 & 1.14 & 35.4 & 34.5 & 33.8 & 34.1 & 21.0 \\
J114422+401221 & 290 & 45 & 0.70 & 34.9 & 34.6 & 33.9 & 34.6 & 20.9 \\
J141454+054047 & 30 & 3.5 & -0.16 & 32.8 & 34.4 & 33.0 & 32.9 & 21.0 \\
J141612+122340 & 400 & 15 & -0.19 & 34.6 & 35.0 & 33.3 & 34.7 & 21.0 \\
J142856+165339 & 330 & 47 & 0.53 & 35.0 & 34.8 & 33.6 & 34.1 & 20.8 \\
J142947+064334 & 360 & 21 & -0.11 & 34.6 & 35.1 & 33.2 & 34.7 & 21.3 \\
J152141+075921 & 280 & 21  & 0.55 & 34.6 & 34.5 & 33.3 & 34.1 & 20.4 \\
J152521+075720 & 350 & 35 & 0.59 & 34.9 & 34.6 & 33.4 & 34.0 & 21.0 \\
J161245+081701 & 450 & 28 & -0.11 & 34.9 & 35.2 & 33.5 & 34.7 & 21.0 \\
J210358-072802 & 500 & 46 & 0.05 & 35.2 & 35.3 & 34.2 & 35.7 & 21. 0\\
Haro 11 & 160 & 45 & 0.10 & 34.7 & 35.2 & 34.3 & 34.9 & 21.3 \\
VV 114  & 130 & 60 & -0.04 & 34.7 & 35.5 & 34.9 & 35.6 & 21.1 \\
NGC 1140 & 34 & 4.8 & 0.76 & 33.0 & 33.6 & 33.5 & 34.0 & 20.9 \\
SBS 0335-052 & 38 & 2.3 & 0.86 & 32.7 & 33.2 & 32.2 & 32.1 & 21.0 \\
Tol 0440-381 & 120 & 33 & 0.82 & 34.4 & 34.4 & 34.2 & 34.7 & 20.9 \\
NGC 1705 & 21 & 1.0 & 0.80 & 32.1 & 32.9 & 32.7 & 33.2 & 20.4 \\
NGC 1741 & 38 & 4.6 & -0.12 & 33.0 & 34.5 & 33.6 & 34.3 & 21.1 \\
I~Zw 18 & $<10$ & $<1$ & $<1.80$ & $ <31.8$ & 31.9 & 32.1 & 31.3 & 21.0\\
NGC 3310 & 180 & 22 & 0.90 & 34.4 & 34.1 & 33.7 & 34.4 & 21.1 \\
Haro 3 & $<10$ & $<1.0$ & $<0.39$ & $<31.8$ & 33.3 & 33.2 & 33.5 & 21.1\\
NGC 3690 & 100 & 12 & -0.52 & 33.9 & 35.3 & 34.3 & 35.7 & 21.3 \\
NGC 4214 & 34 & 1.0 & 0.89 & 32.3 & 32.8 & 32.6 & 33.5 & 21.1 \\
Mrk 54 & 58 & 13 & -0.21 & 33.7 & 35.0 & 34.3 & 35.1 & 20.8 \\
M 83 &  $<10$ & $<6.4$ & $<0.26$ & $<32.6$ & 34.2 & 34.9 & 35.5 & 20.9 \\
NGC 5253 &  17 &  0.6 & 0.26 & 31.8 & 33.2 & 33.2 & 32.9 & 21. 2 \\ 
IRAS 19245-4140 & 65 & 5.4 & 0.41 & 33.3 & 34.0 & 33.1 & 33.6 & 20.7 \\
NGC 7673 & 82 & 30 & 0.80 & 34.2 & 34.4 & 34.3 & 34.7 & 21.0 \\
NGC 7714 & 46 &  4.6 & -0.18 & 33.1 & 34.5 & 33.9 & 34.9 &  21.4 
\enddata
\tablenotetext{a}{The outflow velocity of the strong interstellar absorption-lines arising in the warm ionized gas.  The uncertainties are generally dominated by systematic uncertainties in the fit to the adjacent continuum.  A15 estimate the resulting uncertainties to be $\pm$ 0.05 dex.}
\tablenotetext{b}{The mass outflow rate derived from the outflow velocities, an estimate of the outflow column density, and the column-density-weighted mean radius of the absorbing gas (see text).  A uniform column density of 10$^{21}$ cm$^{-2}$ was assumed, leading to an estimated uncertainty of $\pm$ 0.25 dex (see Figure 4). A uniform radius of 2 $r_*$ was assumed for the calculation.}
\tablenotetext{c}{The momentum flux in the outflow. This is just the mass outflow rate (Column 3) times the outflow velocity.  The uncertainties will be dominated by the former.}
\tablenotetext{d}{The momentum flux from the starburst.  This is assumed to be proportional to the star-formation rate (see text). This conversion may introduce systematic uncertainties that are hard to quantify (Leitherer et al. 1999).}
\tablenotetext{e}{The critical momentum flux needed from the starburst to overcome gravity acting on a cloud with a column density of $10^{21}$ cm$^{-2}$ located at the radius of the starburst. The uncertainties are all systematic, driven by these adopted values.}
\tablenotetext{f}{The effective  column density of dusty gas along the line-of-sight, based on the effective FUV optical depth (Column  10, Table 1), the Calzetti et al. (2000) FUV starburst  attenuation law, and the assumption of a constant dust/gas-phase-metals ratio.  The uncertainties are similar to the observed spread in the distribution ($\pm$ 0.3 dex).}

\end{deluxetable}
\clearpage
\end{landscape}

\end{document}